\documentclass[letterpaper,twocolumn,twoside]{IEEEtran}

\usepackage{graphicx}
\usepackage{float}
\usepackage{amsthm}
\usepackage{amsfonts}
\usepackage{amsmath}
\usepackage{listings}
\usepackage{cite}
\usepackage{psfrag}
\usepackage{pstricks}
\restylefloat{figure}
\theoremstyle{plain} 
\theoremstyle{definition} 
\theoremstyle{plain} 

\newlength{\partitionDiagramWidth}
\newcommand{\Frac}[2]{{{#1}/{#2}}}  % an "inert" form of \frac

\def\argmin{\mathop{\mathrm{arg\,min}}}
\def\E{\mathbb{E}}
\def\R{\mathbb{R}}

\def\half{{\textstyle\frac{1}{2}}}
\def\Abf{\mathbf{A}}
\def\abf{\mathbf{a}}
\def\dbf{\mathbf{d}}
\def\Ibf{\mathbf{I}}
\def\sbf{\mathbf{s}}
\def\wbf{\mathbf{w}}
\def\xbf{\mathbf{x}}
\def\xhat{\widehat{x}}
\def\xbfhat{\widehat{\mathbf{x}}}
\def\ybf{\mathbf{y}}
\def\zbf{\mathbf{z}}
\def\taubf{\boldsymbol{\tau}}
\def\ubf{\mathbf{u}}
\def\taubfhat{\widehat{\boldsymbol{\tau}}}
\def\J{\mathcal{J}}
\def\N{\mathcal{N}}
\def\Op#1{\mathrm{#1}}
\def\Q{\Op{Q}}
\def\arr{\rightarrow}

\begin{document}

\title{Message-Passing Estimation from \\ Quantized Samples%
\thanks{This material is based upon work supported by
  the National Science Foundation under Grant No.\ 0729069 and
  by the DARPA InPho program through the US Army Research Office award
  W911-NF-10-1-0404\@.
  The material in this paper was presented in part at the
  IEEE International Symposium on Information Theory,
  St.\ Petersburg, Russian, July--August 2011.}}
\author{Ulugbek Kamilov, Vivek~K~Goyal, and Sundeep Rangan%
\thanks{U. Kamilov (email: ulugbek.kamilov@epfl.ch) is with
        \'{E}cole Polytechnique F\'{e}d\'{e}rale de Lausanne.
        This work was completed while he was with
        the Research Laboratory of Electronics,
        Massachusetts Institute of Technology.}
\thanks{V. K. Goyal (email: vgoyal@mit.edu) is with
        the Department of Electrical Engineering and Computer Science and
        the Research Laboratory of Electronics,
        Massachusetts Institute of Technology.}
\thanks{S. Rangan (email: srangan@poly.edu) is with the Polytechnic Institute
        of New York University.}%
}

\markboth{Message-Passing Estimation from Quantized Samples}
        {Kamilov, Goyal, and Rangan}

\maketitle

\begin{abstract}
Estimation of a vector from quantized linear measurements is a common
problem for which simple linear techniques are suboptimal---sometimes
greatly so.  This paper develops generalized approximate message passing
(GAMP) algorithms for minimum mean-squared error estimation of a random vector
from quantized linear measurements, notably allowing the linear expansion
to be overcomplete or undercomplete and the scalar quantization to be
regular or non-regular.  GAMP is a recently-developed class of algorithms
that uses Gaussian approximations in belief propagation and allows arbitrary
separable input and output channels.  Scalar quantization of measurements
is incorporated into the output channel formalism, leading to the first
tractable and effective method for high-dimensional estimation problems
involving non-regular scalar quantization.  Non-regular quantization is
empirically demonstrated to greatly improve rate--distortion performance
in some problems with oversampling or with undersampling combined with
a sparsity-inducing prior.
Under the assumption of a Gaussian measurement matrix with i.i.d.\ entries,
the asymptotic error performance of GAMP can be accurately predicted and
tracked through the state evolution formalism.  We additionally use
state evolution to design MSE-optimal scalar quantizers for GAMP
signal reconstruction and empirically demonstrate the superior
error performance of the resulting quantizers.
\end{abstract}

\begin{IEEEkeywords}
analog-to-digital conversion,
approximate message passing,
belief propagation,
compressed sensing,
frames,
non-regular quantizers,
Slepian--Wolf coding,
quantization,
Wyner--Ziv coding
\end{IEEEkeywords}

\section{Introduction}
\label{Sec:Intro}
\IEEEPARstart{E}{stimation} of a signal from quantized samples is a fundamental problem
in signal processing.  It arises both from the discretization
in digital acquisition devices and the quantization performed for lossy compression.

This paper considers of estimation of an i.i.d.\ vector $\xbf$
from quantized transformed samples of the form $\Q(\zbf)$ where
$\zbf=\Abf\xbf$ is a linear transform of $\xbf$ and
$\Q(\cdot)$ is a scalar (componentwise separable) quantization operator.
Due to the transform $\Abf$, the components of $\zbf$ may be
correlated.
Even though the traditional transform
coding paradigm demonstrates the advantages of expressing
the signal with independent components prior to coding~\cite{Goyal:01a}, quantization of vectors with
correlated components nevertheless arises in a range of circumstances.
For example, to model oversampled analog-to-digital conversion (ADC),
we may write a vector of time-domain samples as $\zbf  = \Abf \xbf$,
where the entries of the
vector $\xbf$ are statistically independent Fourier
components and $\Abf$ is an oversampled inverse discrete Fourier transform.
The oversampled ADC quantizes the correlated time-domain samples
$\zbf$, as opposed to the Fourier coefficients $\xbf$.
Distributed sensing also necessitates quantization of components that
are not independent since decorrelating transforms may not be 
possible prior to the quantization.
More recently, compressed sensing has become a motivation to consider
quantization of randomly linearly mixed information,
and several sophisticated reconstruction approaches
have been proposed~\cite{ZymnisBC:10,JacquesHF:11,LaskaBDB:11}.

Estimation of a vector $\xbf$ from quantized samples
of the form $\Q(\Abf\xbf)$ is challenging because the
quantization function $\Q(\cdot)$ is nonlinear and 
the transform $\Abf$ couples, or ``mixes,'' the components 
of $\xbf$, thus necessitating joint estimation.
Although reconstruction from quantized samples
is typically linear, more sophisticated, nonlinear techniques
can offer significant improvements in the case of 
quantized transformed data.
A key example ADC, where the improvement from replacing
conventional linear estimation with nonlinear estimation increases with the
oversampling factor~\cite{ThaoV:94,ThaoV:94b,GoyalVT:98,RanganG:01,Cvetkovic:03,
BenedettoPY:06,BodmannP:07,BodmannL:08,Powell:10}.

This paper focuses on using
a simple message-passing algorithm based on
belief propagation (BP).
Implementation of BP for estimation of a continuous-valued quantity
requires discretization of densities;
this is inherently inexact and leads to high computational complexity.
To handle quantization effects without any heuristic
additive noise model~\cite{Gersho:78} and with low complexity,
we use a recently-developed Gaussian-approximated BP algorithm, called
\emph{generalized approximate message passing} (GAMP)~\cite{Rangan:10arXiv-GAMP}
or \emph{relaxed belief propagation}~\cite{Rangan:10-CISS}, 
which extends earlier methods~\cite{GuoW:06,BayatiM:11} to nonlinear 
output channels.

\subsection{Contributions}
Gaussian approximations of loopy 
BP have previously been shown to be effective in
several other applications~\cite{BoutrosC:02,TanakaO:05,GuoW:06,Rangan:10-CISS,
DonohoMM:09,BayatiM:11};
for our application to estimation from quantized samples,
the extension to general output
channels~\cite{Rangan:10-CISS,Rangan:10arXiv-GAMP} is essential.
Using this extension to nonlinear output channels, we show that
GAMP-based estimation offer several key benefits:
\begin{itemize}
\item \emph{General quantizers:}  
The GAMP algorithm permits essentially arbitrary
quantization functions $\Q(\cdot)$ including
non-uniform and even non-regular quantizers
(i.e.\ quantizers with cells composed of unions
of disjoint intervals) used, for example, in
Wyner--Ziv coding~\cite{WynerZ:76}
and multiple description coding~\cite{Goyal:01b}.
In Section \ref{Sec:Simulations},
we will demonstrate that a non-regular
modulo quantizer can provide performance
improvements for correlated data.  We believe that the GAMP
algorithm provides the first tractable estimation
method that can exploit such quantizers.

\item \emph{General priors:}  GAMP-based estimation can 
incorporate a large class of priors on the components
of $\xbf$, provided that the components are independent.
For example, in Section \ref{Sec:Simulations},
we will demonstrate the algorithm on recovery of 
vectors with sparse priors arising in 
quantized compressed sensing
\cite{ZymnisBC:10,JacquesHF:11,LaskaBDB:11}.

\item \emph{Exact characterization with random transforms:}
In the case of certain large random transforms $\Abf$,
the componentwise performance of GAMP-based estimation
can be precisely predicted by a so-called 
\emph{state evolution} (SE)
analysis reviewed in Section \ref{Sec:StateEvolution}.
From the SE analysis, one can precisely evaluate 
any componentwise performance metric, including for example,
mean-squared error (MSE)\@.  In contrast, works such as 
\cite{ThaoV:94,ThaoV:94b,GoyalVT:98,RanganG:01,Cvetkovic:03,
BenedettoPY:06,BodmannP:07,BodmannL:08,Powell:10}
mentioned above have only obtained bounds or scaling laws.

\item \emph{Performance and optimality:}
Our simulations indicate significantly-improved performance over
traditional methods for estimating from quantized samples
in a range of scenarios.  Moreover, for certain large random
sparse transforms, the SE analysis provides testable
conditions under which the GAMP reconstruction is provably 
optimal \cite{Rangan:10-CISS}.

\item \emph{Computational simplicity:} The GAMP
algorithm is computationally extremely fast.
Our simulation and SE analysis indicate good
performance with a small number
of iterations (10 to 20 in our experience),
with the dominant computational cost per iteration
simply being multiplication by $\Abf$ and $\Abf^T$.

\item \emph{Applications to optimal quantizer design:}
When quantizer outputs are used as inputs to a nonlinear estimation algorithm,
minimizing the MSE between quantizer inputs and outputs
is generally not
equivalent to minimizing the MSE of the final reconstruction~\cite{MisraGV:11}.
To optimize the quantizer for the GAMP algorithm, 
we use the fact that the MSE under large random mixing matrices $\Abf$
can be predicted accurately from a set of simple SE
equations~\cite{GuoW:07,Rangan:10arXiv-GAMP}.  
Then, by modeling the quantizer as a part of the measurement channel,
we use the SE formalism to optimize the quantizer to
minimize the asymptotic distortion after the reconstruction by GAMP\@.
Note that our use of random $\Abf$ is for rigor of the SE formalism;
the effectiveness of GAMP does not depend on this.
\end{itemize}

\subsection{Outline}
The remainder of the paper is organized as follows.
Section~\ref{Sec:Background} provides basic background material on
quantization, compressed sensing, and belief propagation.
Section~\ref{Sec:Intro:QLE} introduces the problem of
estimating a random vector from quantized linear transform coefficients.
It concentrates on
geometric insights for both
the oversampled and undersampled settings.
The main results in this paper apply under a Bayesian formulation
introduced in Section~\ref{Sec:BayesianFormulation}.
Note that this Bayesian formulation does not require sparsity of the
signal nor specify undersampling or oversampling.
The use of generalized approximate message passing to find optimal estimates
under this Bayesian formulation is derived in Section~\ref{Sec:QuantizedRBP}.
Section~\ref{Sec:StateEvolution} describes the use of SE to
predict the performance of GAMP for our problem.
Optimization of quantizers using SE is developed in
Section~\ref{Sec:Optimization},
and experimental results are presented in Section~\ref{Sec:Simulations}.
Section~\ref{Sec:Conclusions} concludes the paper.

\subsection{Notation}
Vectors and matrices will be written in boldface type
($\Abf$, $\xbf$, $\ybf$, \ldots)
to distinguish from scalars written in normal weight ($m$, $n$, \ldots).
Random and non-random quantities (or random variables and their realizations)
are not distinguished typographically since the use of capital letters for
random variables would conflict with the convention of using capital letters
for matrices
(or in the case of quantization, an operator on a vector rather than a scalar).
The probability density function (p.d.f.) of random vector $\xbf$
is denoted $p_\xbf$, and the conditional p.d.f.\ of $\ybf$ given $\xbf$
is denoted $p_{\ybf | \xbf}$.
When these densities are separable and identical across components,
we repeat the previous notations: $p_\xbf$ for the scalar p.d.f.\ and
$p_{\ybf | \xbf}$ for the scalar conditional p.d.f\@.
Writing $x \sim \N(a,b)$ indicates that $x$ is a Gaussian random variable
with mean $a$ and variance $b$.
The resulting p.d.f.\ is written as $p_x(t) = \phi( t \,;\, a,\, b )$.

\section{Background}
\label{Sec:Background}
This section establishes concepts and notations central to the paper.
For a comprehensive tutorial history of quantization,
we recommend~\cite{GrayN:98};
for an introduction to compressed sensing,~\cite{CandesW:08};
and for the basics of
belief propagation,~\cite{Pearl:88,RichardsonU:01,YedidiaFW:03}.

\subsection{Scalar Quantization}
\label{Sec:Intro:Quantization}

A $K$-level scalar quantizer $q: \R \rightarrow \R$
is defined by its \emph{output levels} or \emph{reproduction points}
${\mathcal C} = \{c_i\}_{i=1}^K$ and
\emph{(partition) cells} $\{q^{-1}(c_i)\}_{i=1}^K$.
It can be decomposed into a composition of two mappings
$q = \beta \circ \alpha$ where
$\alpha:\R \rightarrow \{1,\,2,\,\ldots,\,K\}$ is the
\emph{(lossy) encoder} and
$\beta:\{1,\,2,\,\ldots,\,K\} \rightarrow {\mathcal C}$ is the
\emph{decoder}.
The boundaries of the cells are called \emph{decision thresholds}.
One may allow $K=\infty$ to denote that ${\mathcal C}$ is countably infinite.

A quantizer is called \emph{regular} when each cell is a convex set,
i.e., a single interval.
Each cell of a regular scalar quantizer thus has a boundary of one point
(if the cell is unbounded) or two points (if the cell is bounded).
If the input to a quantizer is a continuous random variable, then
the probability of the input being a boundary point is zero.
Thus it suffices to specify the cells of a
$K$-point regular scalar quantizer by its
decision thresholds $\{b_i\}_{i=0}^K$,
with $b_0 = -\infty$ and $b_K = \infty$.
The encoder satisfies
\begin{displaymath}
  \alpha(x) = i \qquad \mbox{for $x \in (b_{i-1},\,b_i)$},
\end{displaymath}
and the output for boundary points can be safely ignored.

The lossy encoder of a non-regular quantizer can be decomposed
into the lossy encoder of a regular quantizer followed by a many-to-one
integer-to-integer mapping.
Suppose $K$-level non-regular scalar quantizer $q'$ has decision thresholds
$\{b'_i\}_{i=0}^{K'}$, and let $\alpha$ be the lossy encoder of a
regular quantizer with these decision thresholds.
Since $q'$ is not regular, $K' > K$.
Let $\alpha':\R \rightarrow \{1,\,2,\,\ldots,\,K\}$
denote the lossy encoder of $q'$.
Then
$\alpha' = \lambda \circ \alpha$, where
\begin{displaymath}
 \lambda:\{1,\,2,\,\ldots,\,K'\} \rightarrow \{1,\,2,\,\ldots,\,K\}
\end{displaymath}
is called a \emph{binning function}, \emph{labeling function},
or \emph{index assignment}.
The binning function is not invertible.

The \emph{distortion} of a quantizer $q$ applied to scalar random
variable $x$ is typically measured by the MSE
\begin{displaymath}
  D = \E[(x - q(x))^2].
\end{displaymath}
A quantizer is called optimal at fixed rate $R = \log_2 K$ when it
minimizes distortion $D$ among all $K$-level quantizers.
To optimize scalar quantizers under MSE distortion,
it suffices to consider only regular quantizers;
a non-regular quantizer will never perform strictly better.

While regular quantizers are optimal for the standard lossy compression problem,
non-regular quantizers are sometimes useful when some information aside from
$q(x)$ is available when estimating $x$.
Two key examples are Wyner--Ziv coding~\cite{WynerZ:76}
and multiple description coding~\cite{Goyal:01b}.
One method for Wyner--Ziv coding is to apply Slepian--Wolf coding 
across a block of samples after regular scalar quantization~\cite{LiuCLX:06};
the Slepian--Wolf coding is binning, but across a block rather than
for a single scalar.
In multiple description scalar quantization~\cite{Vaishampayan:93},
two binning functions are used that together are invertible but
individually are not.
In these uses of non-regular quantizers, side information aids in
recovering $x$ with resolution commensurate with $K'$ while the rate
is only commensurate with $K$, with $K' > K$.

Optimization of a quantizer can rarely be done exactly or analytically.
One standard way of optimizing $q$ is via the \emph{Lloyd algorithm},
which iteratively updates the decision boundaries and output levels by
applying necessary conditions for quantizer optimality.

A quantizer $\Op{Q}: \R^m \rightarrow \R^m$ is called a scalar quantizer
when it is the Cartesian product of $m$ scalar quantizers
$q_i : \R \rightarrow \R$.
In this paper, $\Op{Q}$ always represents a scalar quantizer
with component quantizers $\{q_i\}_{i=1}^m$.

\subsection{Compressed Sensing}
\label{Sec:Intro:CS}
Conventionally, one does not attempt to estimate an $n$-dimensional
signal $\xbf$ from fewer than $n$ scalar quantities;
it would not seem to work from a simple counting of degrees of freedom.
Compressed sensing (CS)~\cite{CandesRT:06-IT,CandesT:06,Donoho:06}
encapsulates a variety of techniques for estimating $\xbf$ from $m < n$
scalar linear measurements, possibly including some noise,
by exploiting knowledge that $\xbf$ is sparse or approximately sparse
in some given transform domain.
Measurements are of the form
\begin{equation}
  \label{Equ:CS:Measurement}
    \zbf = \Abf \xbf,
\end{equation}
where $\Abf \in \R^{m \times n}$ is the \emph{measurement matrix},
or
\begin{equation}
  \label{Equ:CS:MeasurementWithNoise}
    \ybf = \zbf + \dbf = \Abf \xbf + \dbf,
\end{equation}
where $\dbf \in \R^m$ is additive noise.
Many theoretical guarantees for compressed sensing are given with
high probability of success over a random selection of $\Abf$.
Note that it is always assumed that $\Abf$ is available when estimating
$\xbf$ from $\zbf$ or $\ybf$.

In this paper, we simplify notation and expressions by assuming that $\xbf$
itself is sparse or approximately sparse without requiring the use of a
transform domain.
Also, since our focus is on estimation in the presence of degradation
of measurements caused by quantization, we do not consider further the
noiseless measurement model~\eqref{Equ:CS:Measurement}.

The most commonly-studied estimator for the
measurement model~\eqref{Equ:CS:MeasurementWithNoise} is the
\emph{lasso} estimator~\cite{Tibshirani:96}
$$
  \xbfhat = \argmin_{\xbf \in \R^n} \left( \half \| \ybf - \Abf \xbf \|_2^2 + \gamma \| \xbf \|_{1} \right),
$$
where algorithm parameter $\gamma > 0$
trades off data fidelity against sparsity of the solution.
This may be interpreted as a Lagrangian form of the estimator
$$
  \xbfhat = \argmin_{\xbf \, : \, \| \ybf - \Abf \xbf \|_2^2 \leq \epsilon} \| \xbf \|_{1},
$$
which could be justified heuristically by $\| \dbf \|_2^2 \leq \epsilon$.

Most of the CS literature has considered signal recovery with no noise
or with $\| \dbf \|_2^2 \leq \epsilon$.
However, in many practical applications,
measurements have to be discretized to a finite number of bits.
The effect of such quantization on the performance of
CS reconstruction has been studied in~\cite{CandesR:06-DCC,GoyalFR:08}.
In~\cite{SunG:09-ISIT},
high-resolution functional scalar quantization theory was used to
design quantizers for lasso estimation.
Better yet is to change the reconstruction algorithm:
In~\cite{ZymnisBC:10,JacquesHF:11,LaskaBDB:11}, the authors
demonstrate that when $\dbf$ represents quantization error,
$$
  \dbf  =  \Op{Q}(\Abf \xbf) - \Abf \xbf,
$$
significant improvements can be obtained by
replacing the constraint $\| \ybf - \Abf \xbf \|_2^2 \leq \epsilon$
by one that uses the partition cells of the quantizers that compose $\Op{Q}$.

While convex optimization formulations are prominent in CS,
estimation with generic convex program solvers often has
excessively high computational cost.
Thus, there is significant interest in greedy and iterative methods.
The use of belief propagation for CS estimation was first
proposed in~\cite{BaronSB:10};
however, as explained in Section~\ref{Sec:Intro:RBP},
belief propagation has high complexity for the estimation of
continuous-valued quantities.
Lower-complexity approximations to belief propagation were first
proposed for CS estimation in~\cite{DonohoMM:09}.
To handle the effects of quantization precisely,
in this paper we use the generalization of the technique
of~\cite{DonohoMM:09,BayatiM:11} developed by Rangan~\cite{Rangan:10arXiv-GAMP}.

\subsection{Belief Propagation}
\label{Sec:Intro:RBP}

Consider the problem of estimating a random vector $\mathbf{x} \in \R^n$
from noisy measurements $\mathbf{y} \in \R^m$,
where the noise is described by a measurement channel
$p_{\mathbf{y} \mid \mathbf{z}}$
that acts separably and identically on each entry of the vector $\zbf$
obtained via (\ref{Equ:CS:Measurement}).
Moreover suppose that elements in the vector $\mathbf{x}$
are distributed i.i.d.\ according to $p_{\mathbf{x}}$.
We can construct the following conditional probability distribution
over random vector $\mathbf{x}$ given the measurements $\mathbf{y}$:
\begin{equation}
  \label{Equ:RBP:Marginalization}
   p_{\mathbf{x} \mid \mathbf{y}}( \xbf \mid \ybf )
   = \frac{1}{Z} \prod_{j=1}^{n} p_{\mathbf{x}} ( x_j )
   \prod_{i=1}^{m} p_{\mathbf{y} \mid \mathbf{z}} \left( y_i \mid z_i \right),
\end{equation}
where $Z$ is the normalization constant and $z_i = ( \Abf \xbf )_i$.
In principle, it is possible to estimate each $x_j$
by marginalizing this distribution.

Belief propagation replaces the computationally intractable
direct marginalization of $p_{\mathbf{x} \mid \mathbf{y}}$
with an iterative algorithm.
To apply BP, construct a bipartite factor graph
$G = ( V, F, E )$ from (\ref{Equ:RBP:Marginalization})
and pass the following messages along the edges $E$ of the graph:
\begin{subequations}
  \label{Eqs:BP}
\begin{eqnarray}
  \mu_{i \leftarrow j}^{t+1}(x_j)
    & \propto & p_{\mathbf{x}}(x_j)
                \prod_{\ell \neq i} \mu_{\ell \rightarrow j}^{t}(x_j),
                \label{Equ:RBP:varUpdateBp}\\
  \mu_{i \rightarrow j}^{t}(x_j)
    & \propto & \int p_{\mathbf{y} \mid \mathbf{z}} ( y_i \mid z_i )
                  \prod_{k \neq j} \mu_{i \leftarrow k}^{t}(x_j) \,
                  d\xbf_{\setminus j},
                \label{Equ:RBP:mesUpdateBp}
\end{eqnarray}
\end{subequations}
where $\propto$ means that the distribution is to be normalized so that it
has unit integral and integration is over all the elements of $\xbf$ except
$x_j$.
We refer to messages $\{\mu_{i \leftarrow j}\}_{(i,j) \in E}$
as variable updates and to messages
$\{\mu_{i \rightarrow j}\}_{(i,j) \in E}$ as measurement updates.
BP is initialized by setting
$\mu_{i \leftarrow j}^0 (x_j) = p_{\mathbf{x}}(x_j)$.

Earlier works on BP reconstruction have shown that it is
asymptotically MSE optimal under certain verifiable conditions.
These conditions involve simple single-dimensional recursive equations called
\emph{state evolution} (SE), which predicts that BP is optimal when the
corresponding SE admits a unique fixed point~\cite{GuoW:06,GuoW:07}.
However, direct implementation of BP is impractical due to the
dense structure of $\Abf$, which implies that the algorithm must compute the
marginal of a high-dimensional distribution at each measurement node;
i.e., the integration in~\eqref{Equ:RBP:mesUpdateBp} is over many variables.
Furthermore, integration must be approximated through some discrete
quadrature rule.

BP can be simplified through
various Gaussian approximations,
including the \emph{relaxed BP} method~\cite{GuoW:06,Rangan:10-CISS} and 
\emph{approximate message passing (AMP)}~\cite{DonohoMM:09,Rangan:10arXiv-GAMP}.
Recent theoretical work and extensive numerical experiments have demonstrated
that, in the case of certain large random measurement matrices,
the error performance of both relaxed BP and AMP can also be accurately
predicted by SE\@.

\section{Quantized Linear Expansions}
\label{Sec:Intro:QLE}
This paper focuses on the general quantized measurement abstraction of
\begin{equation}
  \label{eq:generalModel}
  \ybf  =  \Op{Q}(\Abf \xbf),
\end{equation}
where $\xbf \in \R^n$ is a signal of interest,
$\Abf \in \R^{m \times n}$ is a linear \emph{mixing matrix},
and $\Op{Q}: \R^m \rightarrow \R^m$ is a scalar quantizer.
We will be primarily interested in (per-component) MSE
$n^{-1} \E[\|\xbf - \xbfhat\|^2]$ for various estimators $\xbfhat$
that depend on $\ybf$, $\Abf$, and $\Op{Q}$.
The cases of $m \geq n$ and $m < n$ are both of interest.
We sometimes use $\zbf = \Abf \xbf$ to simplify expressions.

\subsection{Overcomplete Expansions}
\label{Sec:Intro:QFE}
Let $\Abf \in \R^{m \times n}$ have rank $n$.  Then $\{\abf_i\}_{i=1}^m$ is a
\emph{frame} in $\R^n$, where $\abf_i^T$ is row $i$ of $\Abf$.
%~\cite{DuffinS:52}.
Rank $n$ can occur only with $m \geq n$, so $\Abf \xbf$ is called an
\emph{overcomplete expansion} of $\xbf$ and
$\Op{Q}(\Abf \xbf)$ as in \eqref{eq:generalModel} is called a
\emph{quantized overcomplete expansion}.
In some cases of interest, the frame may be \emph{uniform},
meaning $\| \abf_i \| = 1$ for each $i$, or
\emph{tight}, meaning $\Abf^T \Abf = c\Ibf_n$ for some scalar $c$.

Commonly-used \emph{linear reconstruction} forms estimate
\begin{equation}
  \label{eq:linearEstimate}
  \xbfhat  =  \Abf^\dagger \ybf  =  \Abf^\dagger \Op{Q}(\Abf \xbf),
\end{equation}
where $\Abf^\dagger = (\Abf^T \Abf)^{-1}\Abf^T$ is the pseudoinverse of $\Abf$.
Under several reasonable models, linear reconstruction
has MSE inversely proportional to $m$.
For example, suppose the frame is uniform and tight and
$\xbf$ is an unknown deterministic quantity.
By modeling scalar quantization $y_i = q_i(z_i)$ with an additive noise as
\begin{subequations}
  \label{eqs:additiveNoiseModel}
\begin{equation}
  \label{eqs:additiveNoiseModel_a}
    y_i  =  z_i + d_i
\end{equation}
where
 \begin{eqnarray}
  \label{eqs:additiveNoiseModel_b}
   \E[d_i] & = & 0, \\
  \label{eqs:additiveNoiseModel_c}
   \E[d_i d_j] & = & \sigma_d^2 \delta_{ij},
 \end{eqnarray}
\end{subequations}
one can compute the MSE to be $n\sigma_d^2/m$~\cite{GoyalKK:01}.

\begin{figure*}
 \begin{center}
  \begin{tabular}{ccccc}
     \psfrag{a}[][]{\tiny $\mathcal{S}_1(-2)$}
     \psfrag{b}[][]{\tiny $\mathcal{S}_1(-1)$}
     \psfrag{c}[][]{\tiny $\mathcal{S}_1(0)$}
     \psfrag{d}[][]{\tiny $\mathcal{S}_1(1)$}
     \psfrag{e}[][]{\tiny $\mathcal{S}_1(2)$}
     \psfrag{0}[][]{\scriptsize $\bf 0$}
     \includegraphics[width=\partitionDiagramWidth]{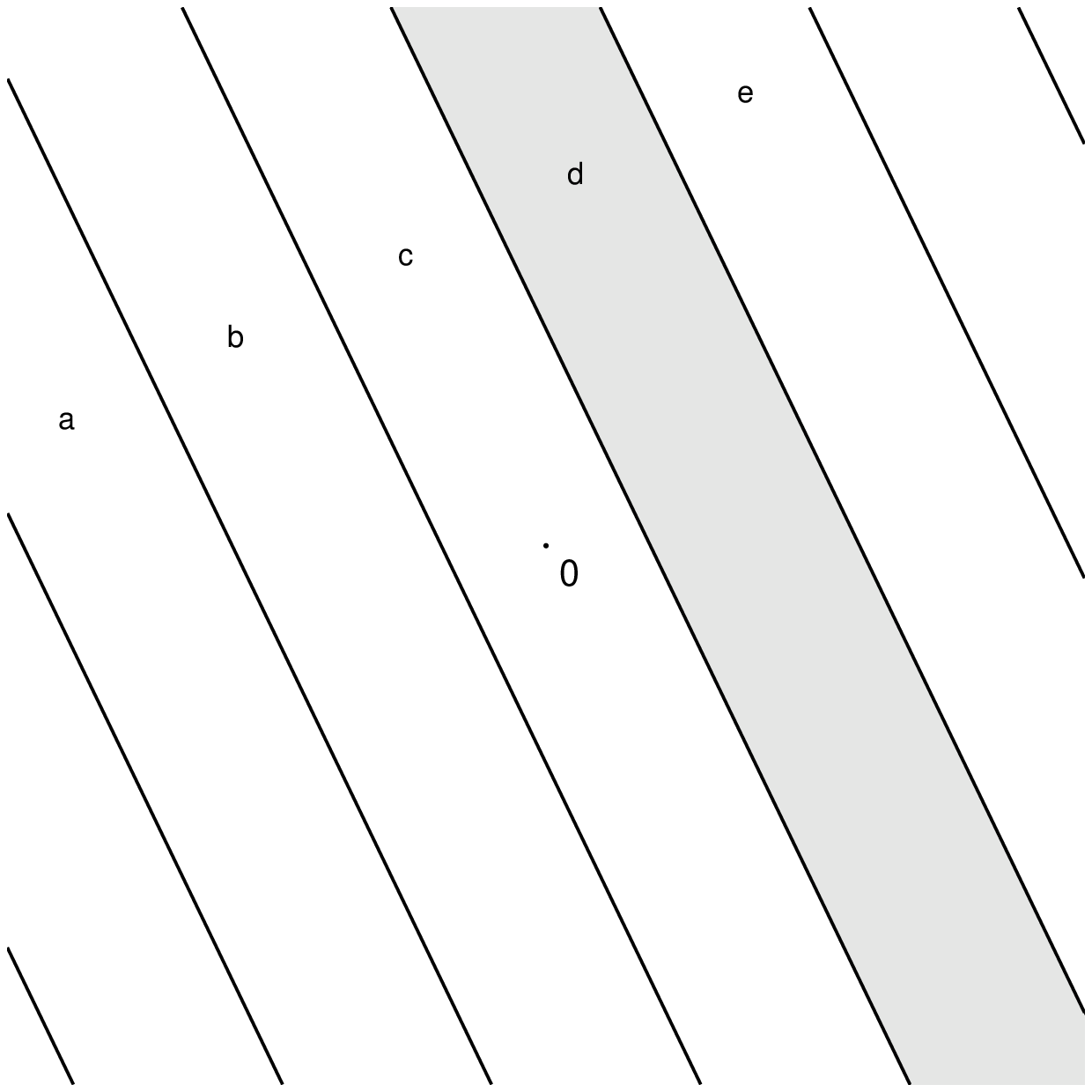} & &
     \psfrag{a}[][]{\tiny $\mathcal{S}_1(1)$}
     \psfrag{b}[][]{\tiny $\mathcal{S}_2(0)$}
     \psfrag{0}[][]{\scriptsize $\bf 0$}
     \includegraphics[width=\partitionDiagramWidth]{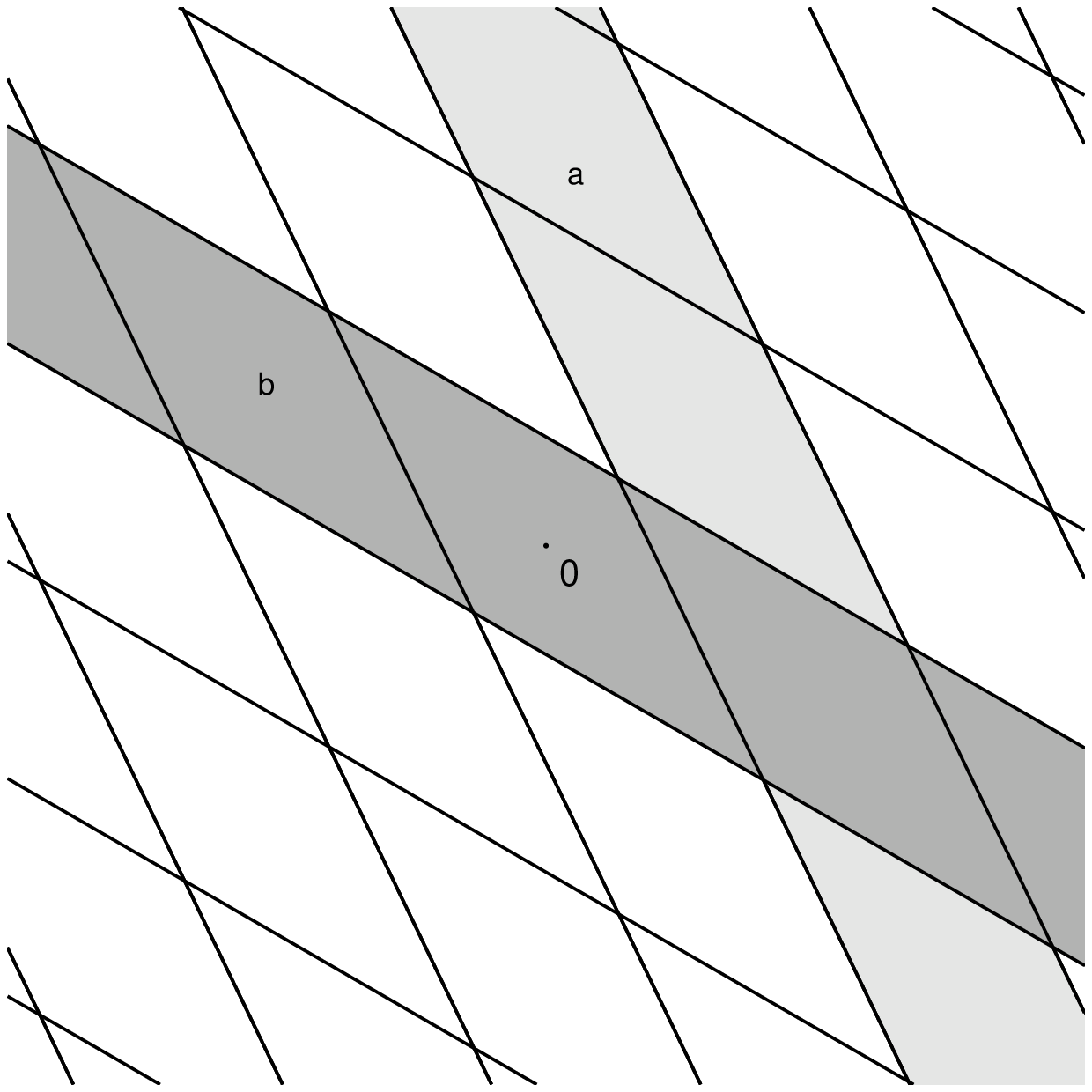} & &
     \includegraphics[width=\partitionDiagramWidth]{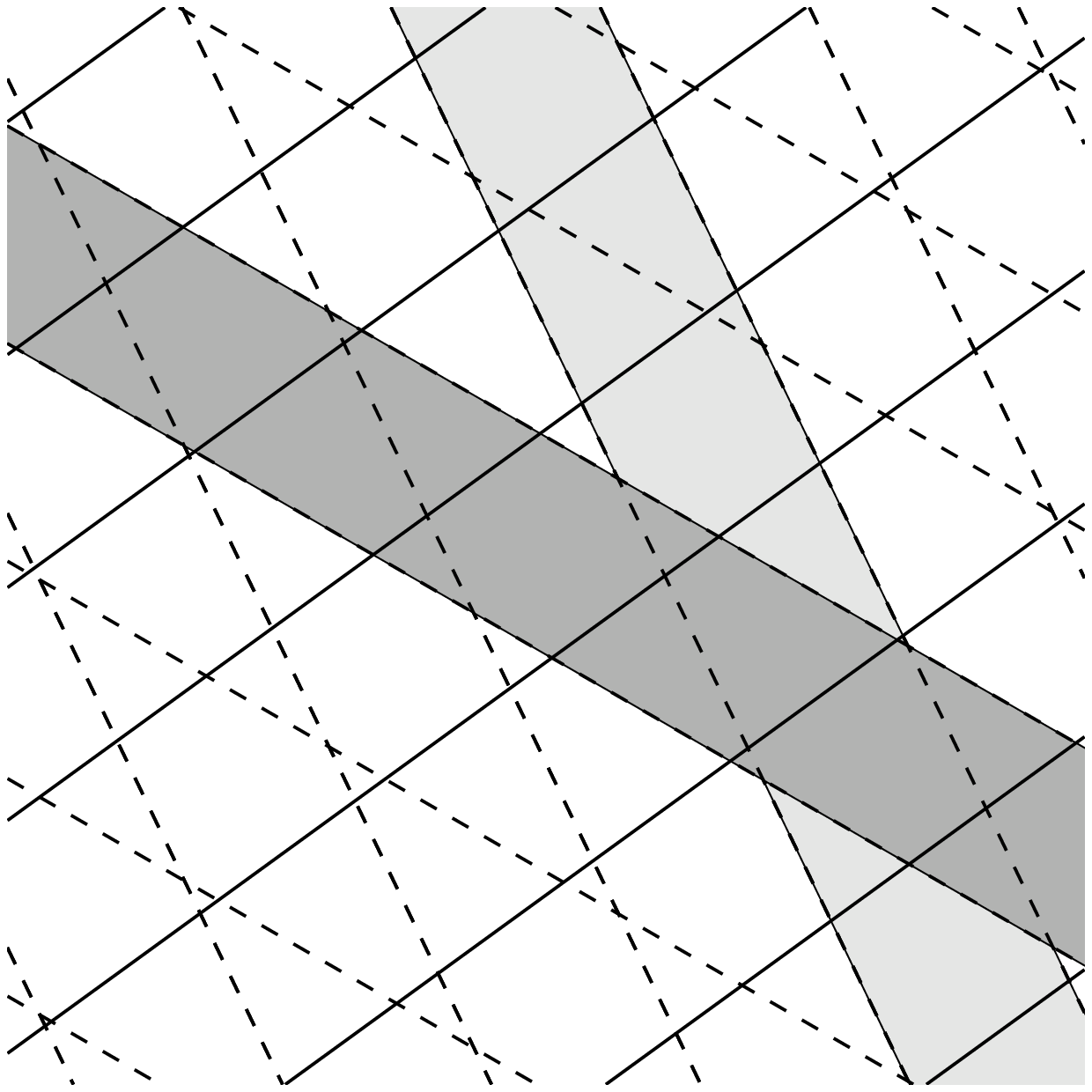} \\
     {\small (a)} & & {\small (b)} & & {\small (c)}
  \end{tabular}
 \end{center}
 \caption{Visualizing the information present in a quantized
       overcomplete expansion of $\xbf \in \R^2$ when each $q_i$ is
       a regular quantizer.
     (a) A single hyperplane wave partition with one single-sample
         consistent set shaded.
     (b) Partition boundaries from two hyperplane waves; $\xbf$ is specified
         to the intersection of two single-sample consistent sets,
         which is a bounded convex cell.
     (c) Partition from part (b) in dashed lines with a third hyperplane wave
         added in solid lines.
     }
 \label{Fig:PartitionRegular}
\end{figure*}

Even when the model \eqref{eqs:additiveNoiseModel}
is accurate~\cite{ViswanathanZ:01},
the linear reconstruction \eqref{eq:linearEstimate}
may be far from optimal.
More sophisticated algorithms have focused on
enforcing \emph{consistency} of an estimate with the
quantized samples.
A nonlinear estimate may exploit the boundedness of the sets
$$
  \mathcal{S}_i(y_i) = \{ \xbf \in \R^n \, | \, q_i(z_i) = y_i \},
\qquad
  i = 1,\,2,\,\ldots,\,m,
$$
which we call \emph{single-sample consistent sets}.
Assuming for now that scalar quantizer $q_i$ is regular and
its cells are bounded,
the boundary of $\mathcal{S}_i(y_i)$ is two parallel hyperplanes.
The full set of hyperplanes obtained for one index $i$ by varying
$y_i$ over the output levels of $q_i$
is called a hyperplane wave partition~\cite{ThaoV:96},
as illustrated for a uniform quantizer in Figure~\ref{Fig:PartitionRegular}(a).
The set enclosed by two neighboring hyperplanes in a
hyperplane wave partition is called a \emph{slab};
one slab is shaded in Figure~\ref{Fig:PartitionRegular}(a).
Intersecting $\mathcal{S}_i(y_i)$ for $n$ distinct indexes specifies
an $n$-dimensional parallelotope as illustrated in
Figure~\ref{Fig:PartitionRegular}(b).
Using more than $n$ of these single-sample consistent sets restricts $\xbf$ to a
finer partition, as illustrated in Figure~\ref{Fig:PartitionRegular}(c)
for $m = 3$.

The intersection
$$\mathcal{S}(\ybf) = \bigcap_{i=1}^m \mathcal{S}_i(y_i)$$
is called the \emph{consistent set}.
Since each $\mathcal{S}_i(y_i)$ is convex,
one may reach $\mathcal{S}(\ybf)$ asymptotically
through a sequence of projections onto $\mathcal{S}_i(y_i)$
using each infinitely often~\cite{ThaoV:94,ThaoV:94b}.

In a variety of settings, nonlinear estimates achieve MSE inversely
proportional to $m^2$,
which is the best possible dependence on $m$~\cite{ThaoV:96}.
The first result of this sort was in~\cite{ThaoV:94}.
When $\Abf$ is an oversampled discrete Fourier transform matrix
and $\Op{Q}$ is a uniform quantizer,
$\zbf = \Abf \xbf$ represents uniformly quantized samples above Nyquist rate
of a periodic bandlimited signal.
For this case, it was proven in~\cite{ThaoV:94} that any
$\xbfhat \in \mathcal{S}(\ybf)$ has $O(m^{-2})$ MSE,
under a mild assumption on $\| \xbf \|$.
This was extended empirically to arbitrary uniform frames in~\cite{GoyalVT:98},
where it was also shown that consistent estimates can be computed through
a linear program.
The techniques of alternating projections and linear programming
suffer from high computational complexity;
yet, since they generally find a corner of the consistent set
(rather than the centroid), the MSE performance is suboptimal.

Full consistency is not necessary for optimal MSE dependence on $m$.
It was shown in~\cite{RanganG:01} that $O(m^{-2})$ MSE is guaranteed
for a simple algorithm that uses each $\mathcal{S}_i(y_i)$ only once,
recursively,
under mild conditions on randomized selection of $\{\abf_i\}_{i=1}^m$.
These results were strengthened and extended to deterministic frames
in~\cite{Powell:10}.

Quantized overcomplete expansions arise naturally in acquisition subsystems
such as ADCs, where $m/n$ represents oversampling factor relative to
Nyquist rate.
In such systems, high oversampling factor may be motivated by a
trade-off between MSE and power consumption or manufacturing cost:
within certain bounds, faster sampling is cheaper than a
higher number of quantization bits per sample~\cite{Walden:99}.
However, high oversampling does not give a good trade-off between
MSE and raw number of bits produced by the acquisition system:
combining the proportionality of bit rate $R$ to number of samples $m$
with the best-case $\Theta(m^{-2})$ MSE, we obtain $\Theta(R^{-2})$ MSE;
this is poor compared to the exponential decrease of MSE with $R$
obtained with scalar quantization of Nyquist-rate samples.

Ordinarily, the bit-rate inefficiency of the raw output is made
irrelevant by recoding, at or near Nyquist rate, soon after
acquisition or within the ADC\@.
An alternative explored in this paper is to combat this bit-rate inefficiency
through the use of non-regular quantization.

\subsection{Non-Regular Quantization}
The bit-rate inefficiency of the raw output with regular quantization
is easily understood with reference to Figure~\ref{Fig:PartitionRegular}(c).
After $y_1$ and $y_2$ are fixed, $\xbf$ is known to lie in the intersection
of the shaded strips.
Only four values of $y_3$ are possible (i.e., the solid hyperplane wave
breaks $\mathcal{S}_1(1) \cap \mathcal{S}_2(0)$ into four cells),
and bits are wasted if this is not exploited in the representation of $y_3$.

\begin{figure*}
 \begin{center}
  \begin{tabular}{ccccc}
     \psfrag{a}[][]{\tiny $\mathcal{S}_1(0)$}
     \psfrag{b}[][]{\tiny $\mathcal{S}_1(1)$}
     \psfrag{c}[][]{\tiny $\mathcal{S}_1(0)$}
     \psfrag{d}[][]{\tiny $\mathcal{S}_1(2)$}
     \psfrag{e}[][]{\tiny $\mathcal{S}_1(1)$}
     \psfrag{0}[][]{\tiny $\bf 0$}
     \includegraphics[width=\partitionDiagramWidth]{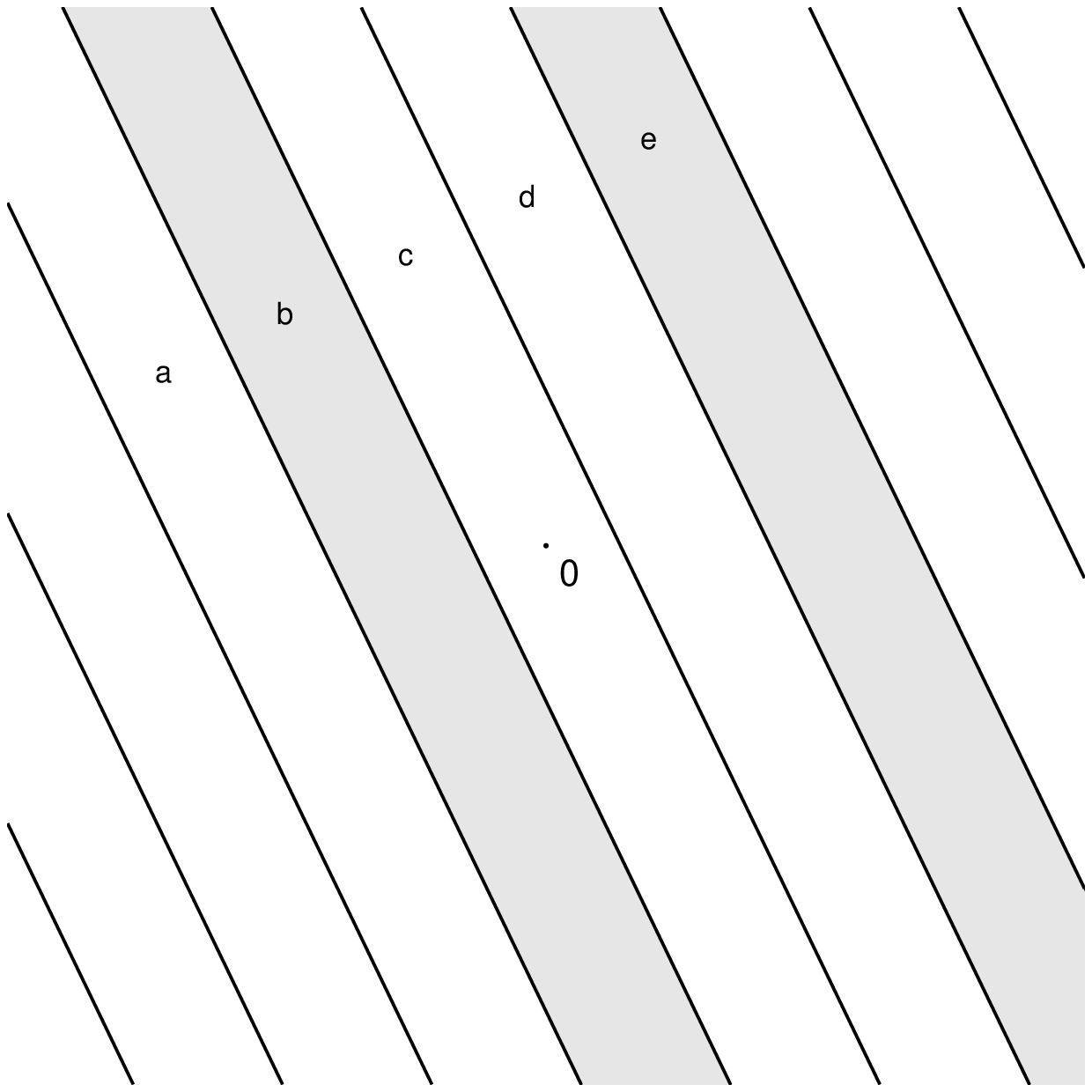} & &
     \includegraphics[width=\partitionDiagramWidth]{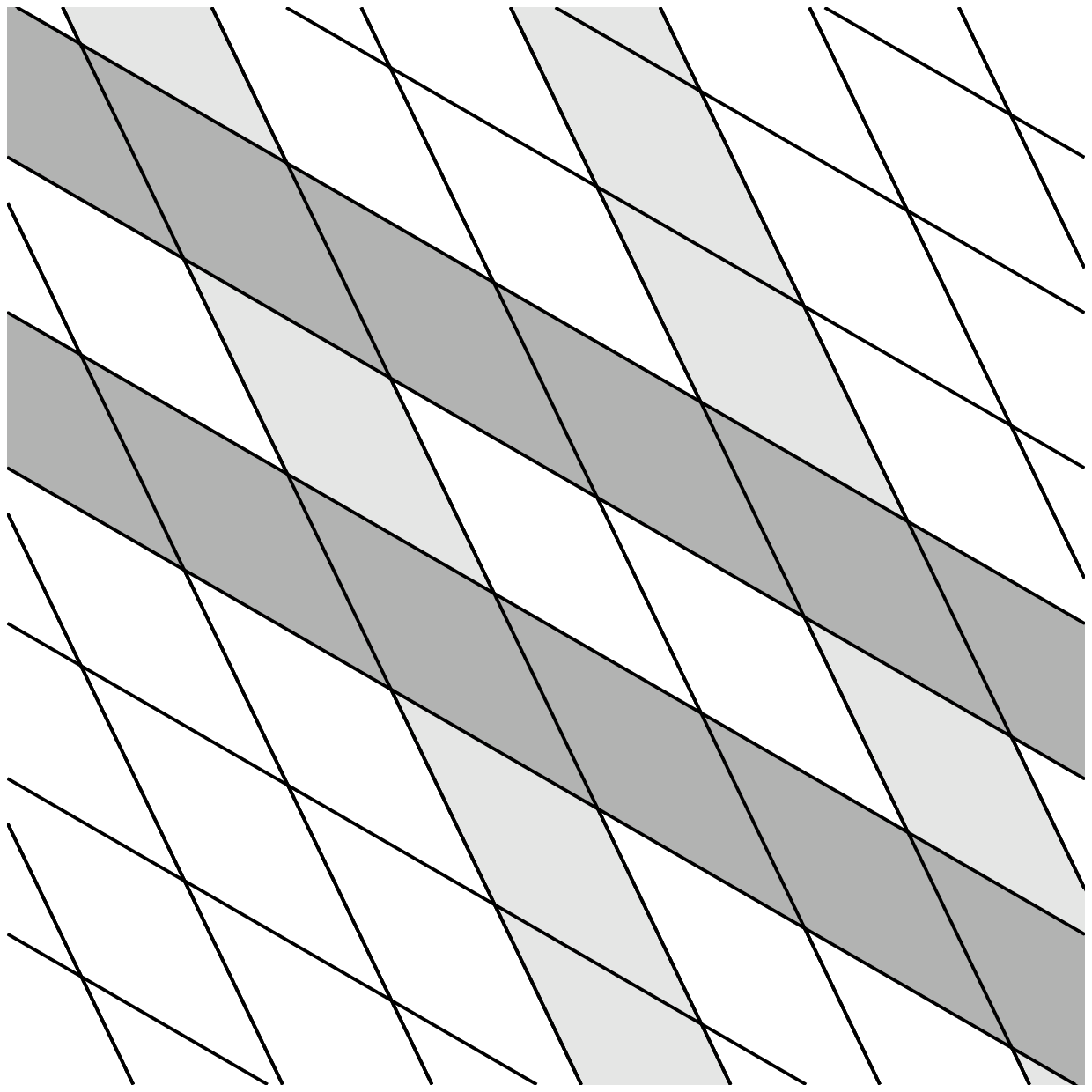} & &
     \psfrag{a}[][]{\scriptsize $\mathcal{S}\,$}
     \includegraphics[width=\partitionDiagramWidth]{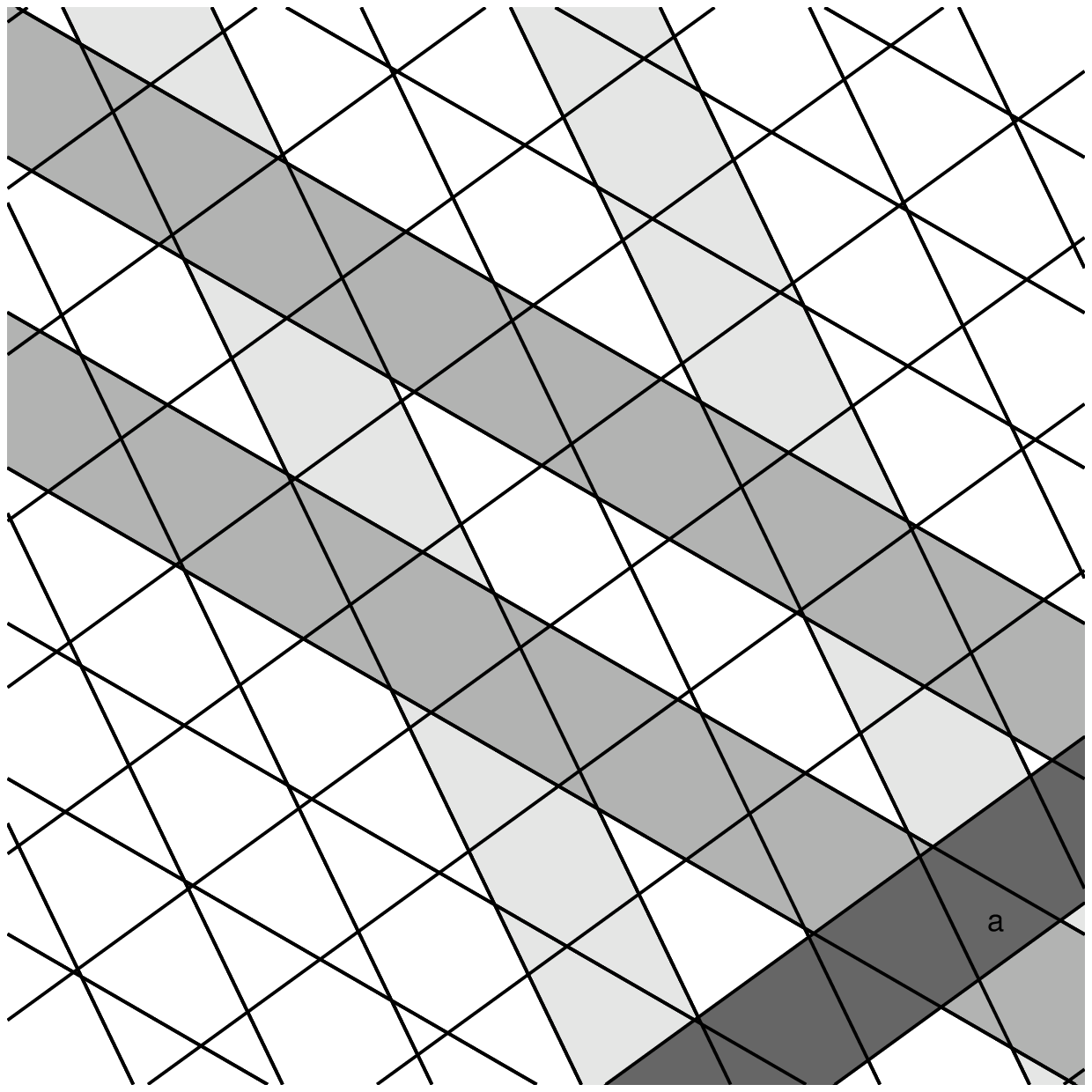} \\
     {\small (a)} & & {\small (b)} & & {\small (c)}
  \end{tabular}
 \end{center}
 \caption{Visualizing the information present in a quantized overcomplete
       expansion of $\xbf \in \R^2$ when using non-regular (binned) quantizers.
     (a) A single hyperplane wave partition with one single-sample
         consistent set shaded.
         Note that binning makes the shaded set not connected.
     (b) Partition boundaries from two hyperplane waves; $\xbf$ is specified
         to the intersection of two single-sample consistent sets, which
         is now the union of four convex cells.
     (c) A third sample now specifies $\xbf$ to within a consistent set
         $\mathcal{S}$ that is convex.
     }
 \label{Fig:PartitionNonregular}
\end{figure*}

Recall the discussion of generating a non-regular quantizer by using
a binning function $\lambda$ in Section~\ref{Sec:Intro:Quantization}.  
Binning does not change the boundaries of the single-sample consistent sets,
but it makes these sets unions of slabs that may not even be connected.
Thus, while binning reduces the quantization rate,
in the absence of side information that specifies which slab contains $\xbf$
(at least with moderately high probability),
it increases distortion significantly.
The increase in distortion is due to \emph{ambiguity} among slabs.
Taking $m > n$ quantized samples together may provide adequate information
to disambiguate among slabs, thus removing the distortion penalty.

The key concepts in the use of non-regular quantization are illustrated
in Figure~\ref{Fig:PartitionNonregular}.
Suppose one quantized sample $y_1$ specifies a single-sample consistent set
$\mathcal{S}_1(y_1)$ composed of two slabs,
such as the shaded region in Figure~\ref{Fig:PartitionNonregular}(a).
A second quantized sample $y_2$ will not disambiguate between the two slabs.
In the example shown in Figure~\ref{Fig:PartitionNonregular}(b),
$\mathcal{S}_2(y_2)$ is composed of two slabs, and
$\mathcal{S}_1(y_1) \cap \mathcal{S}_2(y_2)$ is the union of four
connected sets.  A third quantized sample $y_3$ may now completely
disambiguate; the particular example of $\mathcal{S}_3(y_3)$ shown
in Figure~\ref{Fig:PartitionNonregular}(c) makes
$\mathcal{S} = \mathcal{S}_1(y_1) \cap \mathcal{S}_2(y_2) \cap \mathcal{S}_3(y_3)$
a single convex set.

When the quantized samples together completely disambiguate
the slabs as in the example, the rate reduction from binning comes with
no increase in distortion.
The price to pay comes in complexity of estimation.

The use of binned quantization of linear expansions was introduced
in~\cite{Pai:06}, where the only reconstruction method proposed is
intractable in high dimensions because it is combinatorial over the
binning functions.
Specifically, using the notation from Section~\ref{Sec:Intro:Quantization},
let the quantizer forming $y_i$ be defined by $(\alpha_i,\beta_i,\lambda_i)$.
Then $\lambda_i^{-1}(\beta_i^{-1}(y_i))$ will be a set of possible values of
$\alpha_i(z_i)$ specified by $y_i$.
One can try every combination, i.e., element of
\begin{equation}
\label{eq:binned-combinations}
  \lambda_1^{-1}(\beta_1^{-1}(y_1))
\times
  \lambda_2^{-1}(\beta_2^{-1}(y_2))
\times
  \cdots
\times
  \lambda_m^{-1}(\beta_m^{-1}(y_m)),
\end{equation}
to seek a consistent estimate.
If the binning is effective, most combinations yield an empty consistent set;
if the slabs are disambiguated, exactly one combination yields a non-empty
set, which is then the consistent set $\mathcal{S}$.
This technique has complexity exponential in $m$
(assuming non-trivial binning).
The recent manuscript~\cite{Boufounos:10arXiv} provides bounds on
reconstruction error for consistent estimation with binned quantization;
it does not address algorithms for reconstruction.

This paper provides a tractable and effective method for reconstruction
from a quantized linear expansion with non-regular quantizers.
To the best of our knowledge, this is the first such method.

\begin{figure*}
 \begin{center}
  \begin{tabular}{ccccc}
     \psfrag{a}[][]{}
     \psfrag{b}[][]{}
     \psfrag{c}[][]{}
     \psfrag{d}[][]{}
     \psfrag{e}[][]{}
     \includegraphics[width=\partitionDiagramWidth]{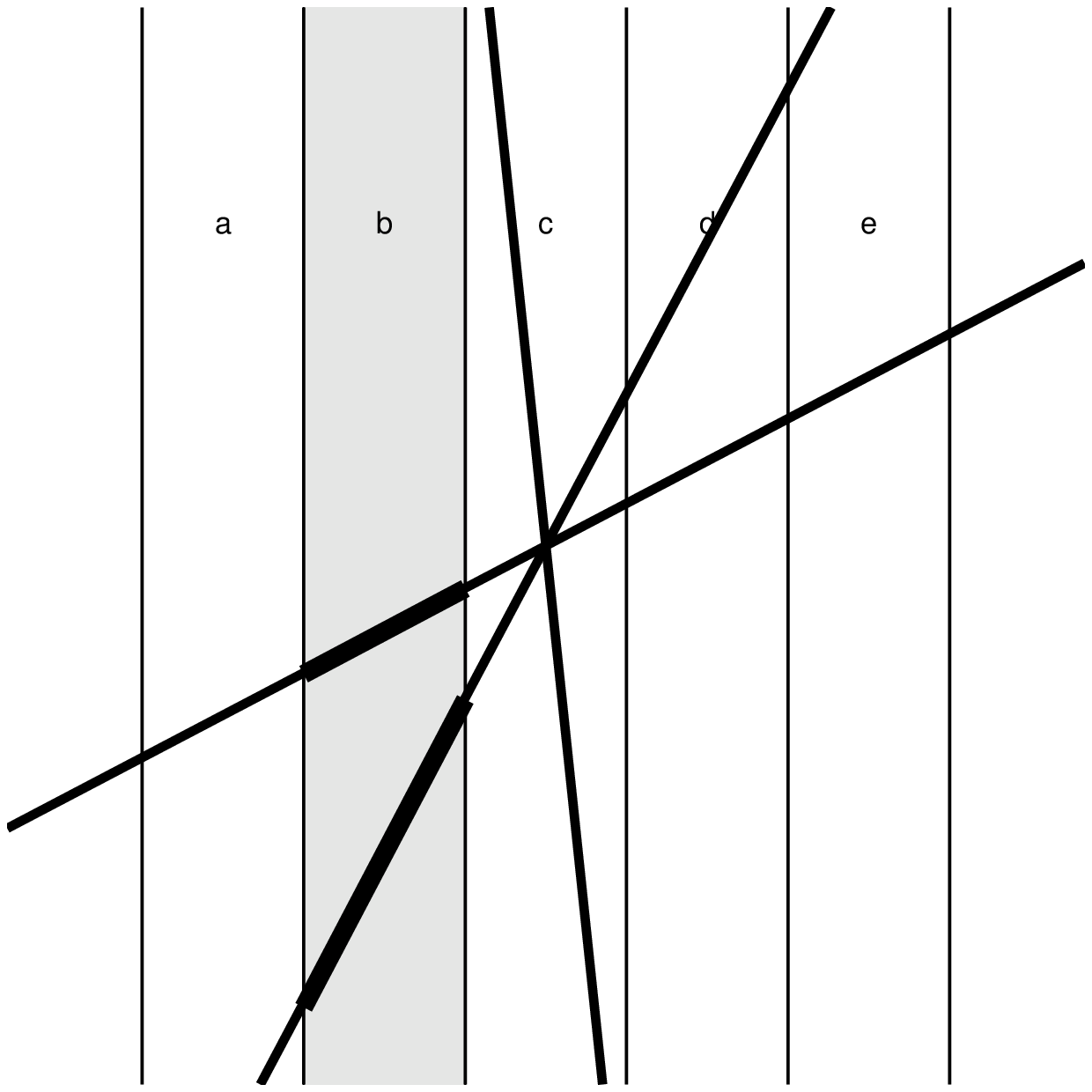} & &
     \includegraphics[width=\partitionDiagramWidth]{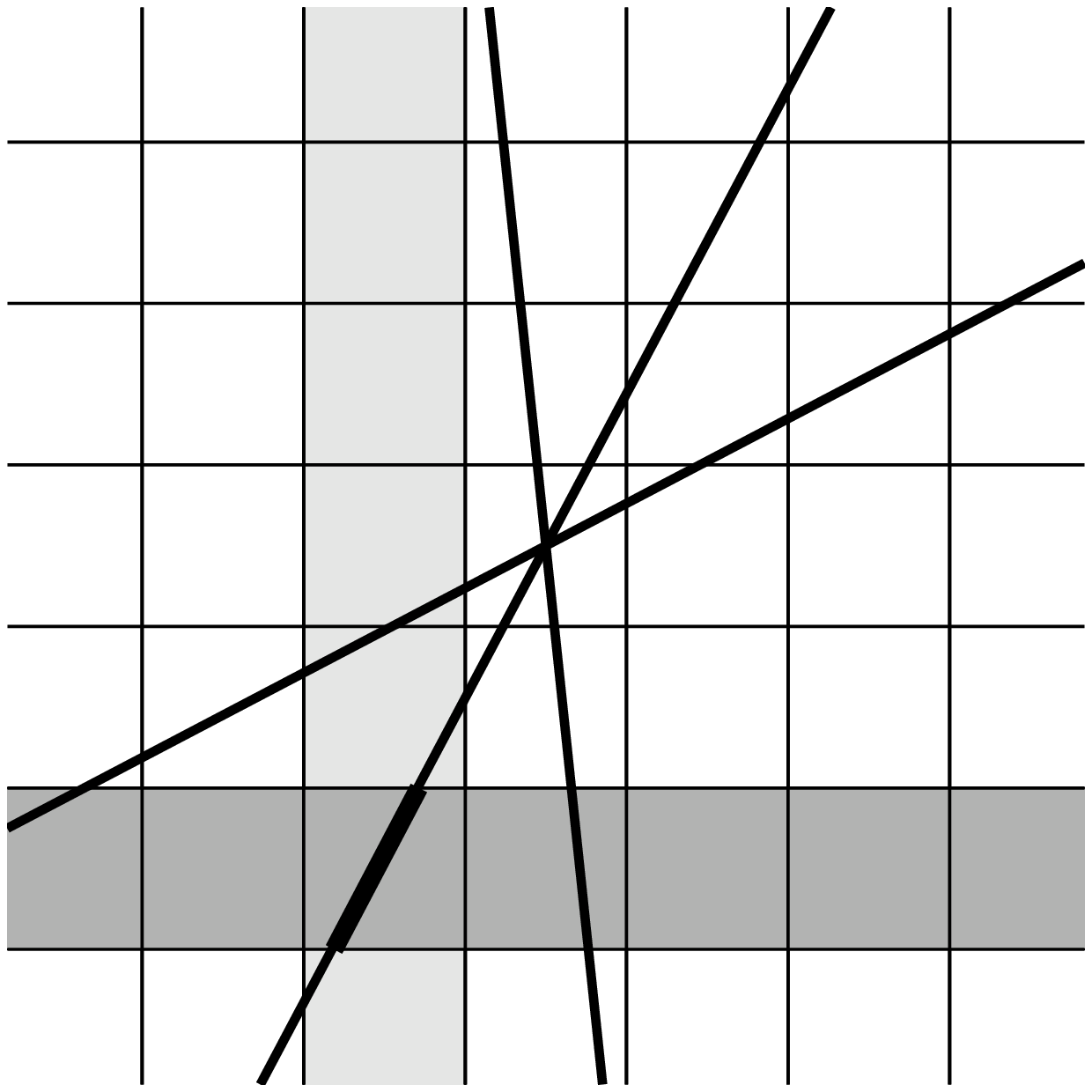} & &
     \psfrag{a}[][]{\scriptsize $\mathcal{S}\,$}
     \includegraphics[width=\partitionDiagramWidth]{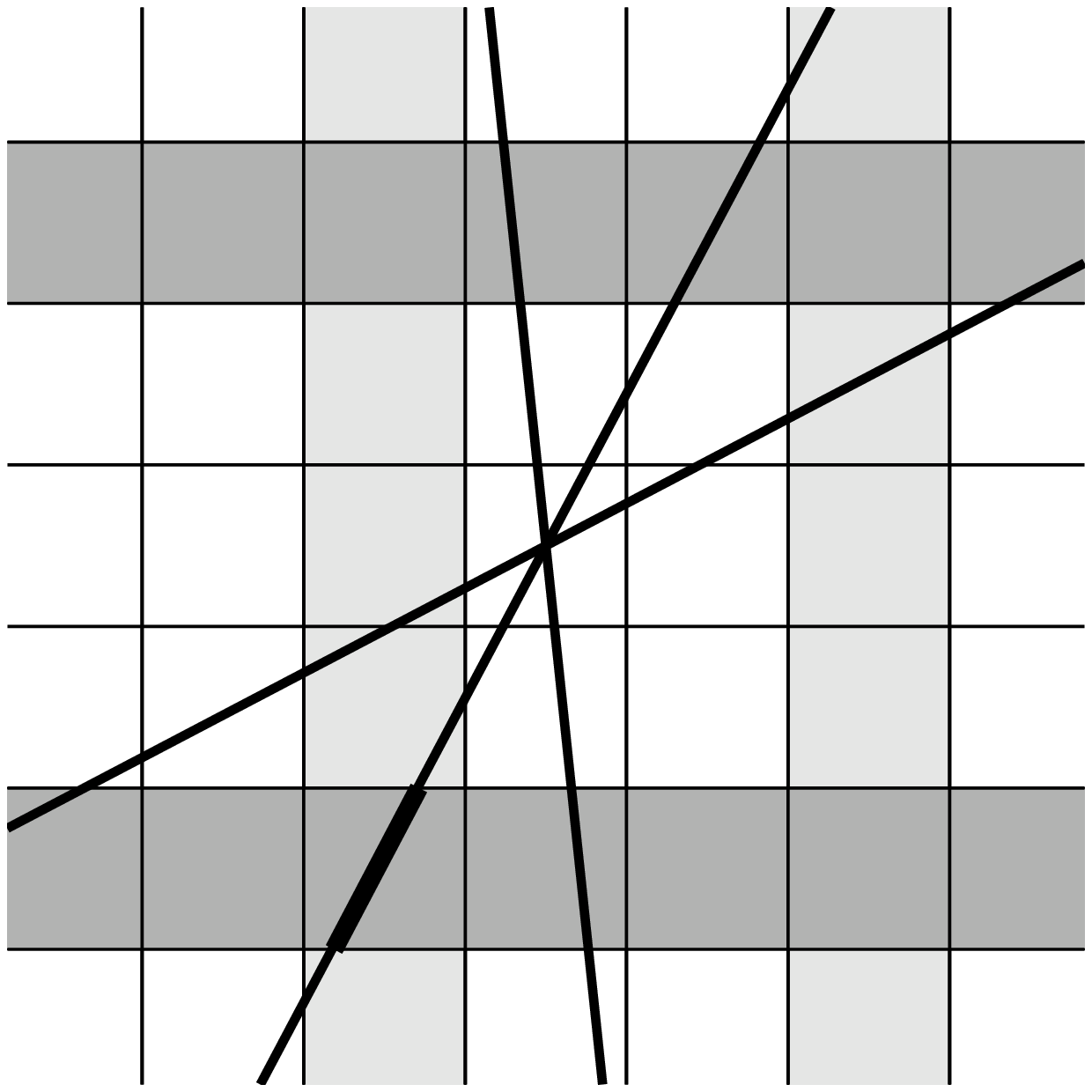} \\
     {\small (a)} & & {\small (b)} & & {\small (c)}
  \end{tabular}
 \end{center}
 \caption{Visualizing the information present in a quantized undercomplete
       expansion $\Op{Q}(\Abf \xbf)$ of a 1-sparse signal $\xbf \in \R^3$ when
       $\Abf \xbf \in \R^2$.  The depicted 2-dimensional plane represents the
       vector of measurements $\zbf = \Abf \xbf$.  Since $\xbf$ is 1-sparse,
       the measurement lies in a union of 1-dimensional subspaces
       (the angled solid lines); since $\xbf$ is 3 dimensional,
       there are three such subspaces.
     (a) Scalar quantization of $z_1$ divides the plane of possible
       values for $\zbf$ into vertical strips.  One particular value of
       $y_1 = q_1(z_1)$ does not specify which entry of $\xbf$
       is nonzero since the shaded strip intersects all the angled solid lines.
       For each possible support, the value of the nonzero entry is specified
       to an interval.
     (b) Scalar quantization of both components of $\zbf$ specifies
       $\zbf$ to a rectangular cell.  In most cases, including the one
       highlighted, the quantized values specify which entry of $\xbf$
       is nonzero because only one angled solid line intersects the cell.
       The value of the nonzero entry is specified to an interval.
     (c) In many cases, including the one highlighted, the quantizers can be
       non-regular (binned) and yet still uniquely specify which entry of $\xbf$
       is nonzero.
     }
 \label{Fig:PartitionUndercomplete}
\end{figure*}

\subsection{Undercomplete Expansions}
Maintaining the quantized measurement model \eqref{eq:generalModel},
let us turn to the case of $m < n$.  We now call $\Op{Q}(\Abf \xbf)$ a
\emph{quantized undercomplete expansion} of $\xbf$.

Since the rank of $\Abf$ is less than $n$, $\Abf$ is a many-to-one mapping.
Thus, even without quantization, one cannot recover $\xbf$ from $\Abf \xbf$.
Rather, $\Abf \xbf$ specifies a proper subspace of $\R^n$ containing $\xbf$;
when $\Abf$ is in general position, the subspace is of dimension $n-m$.
Quantization increases the ambiguity in the value of $\xbf$,
yielding consist sets similar to those depicted in
Figures~\ref{Fig:PartitionRegular}(a) and~\ref{Fig:PartitionNonregular}(a).
However, as described in Section~\ref{Sec:Intro:CS},
knowledge that $\xbf$ is sparse or approximately sparse could be exploited
to enable accurate estimation of $\xbf$ from $\Op{Q}(\Abf \xbf)$.

For ease of explanation,
consider only the case where $\xbf$ is known to be $k$-sparse with $k < m$.
Let $\J \subset \{1,\,2,\,\ldots,\,n\}$ be the support (sparsity pattern)
of $\xbf$, with $|\J| = k$.
The product $\Abf \xbf$ is equal to $\Abf_\J \xbf_\J$,
where $\xbf_\J$ denotes the restriction of the domain of $\xbf$ to $\J$
and $\Abf_\J$ is the $m \times k$ submatrix of $\Abf$ containing the
$\J$-indexed columns.
Assuming $\Abf_\J$ has rank $k$ (i.e., full rank),
$\Op{Q}(\Abf \xbf) = \Op{Q}(\Abf_\J \xbf_\J)$
is a quantized \emph{overcomplete} expansion of $\xbf_\J$.
All discussion of estimation of $\xbf_\J$ from the previous subsections
thus applies, assuming $\J$ is known.

The key remaining issue is that $\Op{Q}(\Abf \xbf)$ may or may not
provide enough information to infer $\J$.
In an overcomplete representation, most vectors of quantizer outputs
cannot occur; this redundancy was used to enable binning in
Figure~\ref{Fig:PartitionNonregular}, and it can be used to
show that certain subsets $\J$ are inconsistent with the sparse
signal model.
In principle, one may enumerate the sets $\J$ of size $k$ and apply a
consistent reconstruction method for each $\J$.
If only one candidate $\J$ yields a non-empty consistent set, then $\J$
is determined.  This is intractable except for small problem sizes
because there are ${n \choose k}$ candidates for $\J$.

The key concepts are illustrated in Figure~\ref{Fig:PartitionUndercomplete}.
To have an interpretable diagram with $k < m < n$,
we let $(k,m,n)=(1,2,3)$ and draw the space of unquantized measurements
$\zbf \in \R^2$.  (This contrasts with Figures~\ref{Fig:PartitionRegular}
and~\ref{Fig:PartitionNonregular} where the space of $\xbf \in \R^2$ is drawn.)
The vector $\xbf$ has one of ${n \choose k} = {3 \choose 1} = 3$ possible
supports $\J$.  Thus, $\zbf$ lies in one of 3 subspaces of dimension 1,
which are depicted by the angled solid lines.
Scalar quantization of $\zbf$ corresponds to separable partitioning of
$\R^2$ with cell boundaries aligned with coordinate axes, as shown with
lighter solid lines.

Only one quantized measurement $y_1$ is not adequate to specify $\J$,
as shown in Figure~\ref{Fig:PartitionUndercomplete}(a)
by the fact that a single shaded cell intersects all the subspaces.%
\footnote{Intersections with two subspaces are shown within the range
of the diagram.}
Two quantized measurements together will usually specify $\J$,
as shown in Figure~\ref{Fig:PartitionUndercomplete}(b) by the fact that
only one subspace intersects the specified square cell;
for fixed scalar quantizers, ambiguity becomes less likely as $k$ decreases,
$n$ increases, $m$ increases, or $\| x \|$ increases.
Figure~\ref{Fig:PartitionUndercomplete}(c) shows a case where non-regular
(binned) quantization still allows unambiguous determination of $\J$.

The na{\"i}ve reconstruction method implied by
Figure~\ref{Fig:PartitionUndercomplete}(c) is to search combinatorially
over both $\J$ and the combinations in (\ref{eq:binned-combinations});
this is extremely complex.
While the use of binning for quantized undercomplete expansions
of sparse signals has appeared in the literature,
first in~\cite{Pai:06} and later in~\cite{Boufounos:10arXiv},
to the best of our knowledge this paper is the first to provide a
tractable and effective reconstruction method.

\section{Estimation from Quantized Samples: Bayesian~Formulation}
\label{Sec:BayesianFormulation}
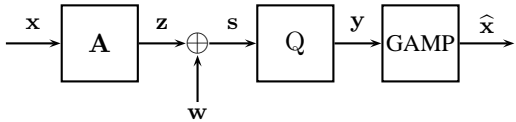
\begin{figure}
  \begin{center}
    \psset{unit=5.2mm}
    \begin{pspicture}(-1.4,2)(11.6,4.5)
      \rput(-0.7,4.5){\small $\xbf$}
      \rput(2.6,4.5){\small $\zbf$}
      \rput(3.5,2.2){\small $\wbf$}
      \rput(4.4,4.5){\small $\sbf$}
      \rput(7.6,4.5){\small $\ybf$}
      \rput(10.9,4.5){\small $\xbfhat$}

      \psline{->}(-1.4,4.0)(0.0,4.0)
      \psframe(0.0,3.0)(2.0,5.0)
      \rput(1.0,4.0){$\Abf$}

      \psline{->}(2.0,4.0)(3.2,4.0)
      \rput(3.5,4.0){\large $\oplus$}
      \psline{->}(3.5,2.5)(3.5,3.7)

      \psline{->}(3.8,4.0)(5.0,4.0)
      \psframe(5.0,3.0)(7.0,5.0)
      \rput(6.0,4.0){$\Q$}

      \psline{->}(7.0,4.0)(8.2,4.0)
      \psframe(8.2,3.0)(10.2,5.0)
      \rput(9.2,4.0){\small GAMP}

      \psline{->}(10.2,4.0)(11.6,4.0)
    \end{pspicture}
  \end{center}
  \caption{Quantized linear measurement model for which GAMP estimator is
    derived.  Vector $\xbf \in \R^n$ with an i.i.d.\ prior is estimated
    from scalar quantized measurements $\ybf \in \R^m$.
    The quantizer input $\sbf$ is the sum of $\zbf = \Abf \xbf \in \R^m$
    and an i.i.d.\ Gaussian noise vector $\wbf$.
    Including noise variance $\sigma^2$ in the model clarifies certain
    derivations; setting the noise variance to zero recovers acquisition model
    (\ref{eq:generalModel}).}
  \label{Fig:ProbMod}
\end{figure}

We now specify more explicitly the class of problems for which we
derive new estimation algorithms.
Generalizing (\ref{eq:generalModel}), let
\begin{equation}
 \label{eq:modelWithNoise}
  \ybf = \Op{Q}(\zbf + \wbf)
\qquad
\mbox{where}
\qquad
  \zbf = \Abf \xbf,
\end{equation}
as depicted in Figure~\ref{Fig:ProbMod}.
The input vector $\xbf \in \R^n$ is random with i.i.d.\ entries
with prior p.d.f.\ $p_\xbf$.
The linear mixing matrix $\Abf \in \R^{m \times n}$ is random with
i.i.d.\ entries $a_{ij} \sim \N(0,1/m)$.
The (pre-quantization) additive noise $\wbf \in \R^m$ is random with
i.i.d.\ entries $w_{i} \sim \N(0,\sigma^2)$.
The quantizer $\Op{Q}$ is a scalar quantizer, and each of its component
quantizers $q_i$ is identical and has $K$ output levels.

The estimator $\xbfhat$ is a function of $\Abf$, $\ybf$, $\Op{Q}$, and $\sigma^2$.
We wish to minimize the MSE $n^{-1}\E[\|\xbf - \xbfhat\|^2]$.

Our primary interest is in the case of $\sigma^2 = 0$, but allowing
a nontrivial distribution for $\wbf$ is not only more general but also
makes the derivations more clear.

\section{Generalized Approximate Message Passing for a Quantizer Output Channel}
\label{Sec:QuantizedRBP}
The acquisition model (\ref{eq:modelWithNoise}) is suitable for GAMP
estimation under the conditions in~\cite{Rangan:10arXiv-GAMP} after one
simple observation:
the mapping from $\zbf$ to $\ybf$ is a separable probabilistic mapping
with identical marginals.
Specifically, quantized measurement $y_i$
indicates $s_i \in q_i^{-1}(y_i)$,
so each component \emph{output channel} can be characterized as
\begin{equation*}
   p_{\mathbf{y} \mid \mathbf{z}} ( y \mid z )
    = \int_{q_i^{-1}(y)} \phi \left( t \,;\, z,\, \sigma^2 \right) \, dt,
\end{equation*}
where $\phi$ is the Gaussian function
\begin{equation*}
   \phi \left( t \,;\, a,\, b \right) = \frac{1}{\sqrt{2\pi b}} \exp \left( -\frac{(t-a)^2}{2b} \right).
\end{equation*}

GAMP can be derived by approximating the updates in
(\ref{Eqs:BP}) by two scalar parameters each and introducing some first-order approximations, as discussed in~\cite{Rangan:10arXiv-GAMP}.
Then given the estimation functions $F_{\textrm{in}}$,
$\mathcal{E}_{\textrm{in}}$, $D_1$, and $D_2$ described below,
for each iteration $t = 0,\, 1,\, 2,\, \dots$,
the GAMP algorithm produces estimates $\xbfhat^{t}$
of the true signal $\xbf$ according to the following rules:
\begin{subequations}
  \label{Eqs:GAMP}
\begin{eqnarray}
  \xbfhat^{t+1} & \equiv & F_{\textrm{in}} \left(\xbfhat^{t}+\frac{ \Abf^T \ubf^{t}}{\left( \Abf^T \right)^2 \taubf^t}, \frac{1}{\left( \Abf^T  \right)^2 \taubf^t}\right),\label{Equ:AMP:varUpdateAmpMean}\\
  \taubfhat^{t+1} & \equiv & \mathcal{E}_{\textrm{in}} \left( \xbfhat^{t}+\frac{ \Abf^T \ubf^{t}}{\left( \Abf^T \right)^2 \taubf^t}, \frac{1}{\left( \Abf^T  \right)^2 \taubf^t} \right),\label{Equ:AMP:varUpdateAmpVar}\\
  \ubf^{t} & \equiv & D_1 \left( \ybf, \Abf \xbfhat^{t} - \ubf^{t-1} \Abf^{2}\taubfhat^{t}, \Abf^2 \taubfhat^{t} + \sigma^2\Ibf_n \right),\label{Equ:AMP:mesUpdateAmpMean}\\
  \taubf^{t} & \equiv & D_2 \left(  \ybf, \Abf \xbfhat^{t} - \ubf^{t-1} \Abf^{2}\taubfhat^{t}, \Abf^2 \taubfhat^{t} + \sigma^2\Ibf_n  \right).\qquad\label{Equ:AMP:mesUpdateAmpVar}
\end{eqnarray}
\end{subequations}
Note that in (\ref{Eqs:GAMP}) the notation $\Abf^2$
denotes the element-wise product of a matrix with itself,
i.e. $(\Abf^2)_{ij} = (\Abf_{ij})^2$.
The estimation functions $F_{\textrm{in}}$, $\mathcal{E}_{\textrm{in}}$,
$D_1$, and $D_2$ described below are applied to their inputs
component-by-component.

We refer to messages
$\{\hat{x}_{j}, \hat{\tau}_{j} \}_{j \in V}$
as variable updates and to messages
$\{u_{i}, \tau_{i} \}_{i \in F}$
as measurement updates.
The algorithm is initialized by setting
$\hat{x}^{0}_{j} = \hat{x}_{\textrm{init}}$,
$\hat{\tau}^{0}_{j} = \hat{\tau}_{\textrm{init}}$, and $u_{i}^{-1} = 0$, where
$\hat{x}_{\textrm{init}}$ and $\hat{\tau}_{\textrm{init}}$
are the mean and variance of the prior $p_{\mathbf{x}}$. 
The nonlinear functions $F_{\textrm{in}}$ and $\mathcal{E}_{\textrm{in}}$
are the conditional mean and variance
\begin{eqnarray*}
  F_{\textrm{in}} \left( q, \nu \right)
    & \equiv & \E \left[ x \mid q \right],\\
  \mathcal{E}_{\textrm{in}} \left( q, \nu \right)
    & \equiv & \mathrm{var} \left( x \mid q \right),
\end{eqnarray*}
where $q = x+v$ with $x \sim p_\xbf$ and $v \sim \N(0,\nu)$.
Note that these functions can easily be
evaluated numerically for any given values of $q$ and $\sigma^2$.
Similarly, the functions $D_1$ and $D_2$ can be computed via
\begin{subequations}
\begin{eqnarray}
  D_1\left(y, \hat{z}, \nu\right)
    & \equiv & \frac{1}{\nu}\left(F_{\textrm{out}}\left(y, \hat{z}, \nu\right) - \hat{z}\right),\\
  D_2\left(y, \hat{z}, \nu\right)
    & \equiv & \frac{1}{\nu}\left( 1 - \frac{\mathcal{E}_{\textrm{out}}\left(y, \hat{z}, \nu\right)}{\nu} \right),\label{Equ:RBP:D2}
\end{eqnarray}
\end{subequations}
where the functions $F_{\textrm{out}}$ and $\mathcal{E}_{\textrm{out}}$
are the conditional mean and variance
\begin{subequations}
\begin{eqnarray}
  F_{\textrm{out}}\left(y, \hat{z}, \nu\right)
    & \equiv & \E \left[ z \mid z \in q_i^{-1} \left(y\right) \right],\\
  \mathcal{E}_{\textrm{out}}\left(y, \hat{z}, \nu\right)
    & \equiv & \mathrm{var} \left( z \mid z \in q_i^{-1} \left(y\right) \right), \label{Equ:RBP:Eout}
\end{eqnarray}
\end{subequations}of the random variable
$z \sim \mathcal{N} \left( \hat{z}, \nu \right)$.
These functions admit closed-form expressions in terms of
$\mathrm{erf}\left(z\right) = \frac{2}{\sqrt{\pi}}\int_{0}^{z}e^{-t^2} \, dt$.

\section{State Evolution for GAMP}
\label{Sec:StateEvolution}
The equations (\ref{Eqs:GAMP})
are easy to implement, however they provide us no insight into the performance
of the algorithm.
The goal of SE equations is to describe the asymptotic behavior of GAMP
under large random measurement matrices $\Abf$.

The SE for our setting in Figure~\ref{Fig:ProbMod} is given by the recursion
\begin{equation}
  \label{Equ:SE:SERecursion}
    \bar{\tau}_{t+1} = \bar{\mathcal{E}}_{\textrm{in}} \left( \frac{1}{\bar{D}_2 \left( \beta\bar{\tau}_{t}, \sigma^2 \right)} \right),
\end{equation}
where $t \geq 0$ is the iteration number, $\beta = \Frac{n}{m}$
is a fixed number denoting the measurement ratio, and $\sigma^2$
is the variance of the additive white Gaussian noise (AWGN),
which is also fixed.
We initialize the recursion by setting
$\bar{\tau}_0 = \hat{\tau}_{\textrm{init}}$,
where ${\tau}_{\textrm{init}}$ is the variance of $x_j$
according to the prior $p_{\mathbf{x}}$.
We define the function $\bar{\mathcal{E}}_{\textrm{in}}$ as
\begin{equation}
   \bar{\mathcal{E}}_{\textrm{in}} \left( \nu \right) = \E \left[ \mathcal{E}_{\textrm{in}} \left( q, \nu \right) \right],
\end{equation}
where the expectation is taken over the scalar random variable
$q = x + v$, with
$x \sim p_{\mathbf{x}}$ and
$v \sim \mathcal{N} ( 0, \nu )$.
Similarly, the function $\bar{D}_2$ is defined as
\begin{equation}
   \bar{D}_2 \left( \nu, \sigma^2 \right) = \E \left[ D_2 \left( y, \hat{z}, \nu + \sigma^2 \right) \right],
\end{equation}
where $D_2$ is given by (\ref{Equ:RBP:D2}) and the expectation is taken over
$p_{\mathbf{y} \mid \mathbf{z}}$ and
$( z, \hat{z} ) \sim \mathcal{N} ( 0, P_z( \nu) )$,
with the covariance matrix
\begin{equation}
   P_z \left( \nu \right) = \left(
      \begin{array}{cc}
	\beta{\hat{\tau}}_{\textrm{init}} & \beta{\hat{\tau}}_{\textrm{init}} - \nu\\
	\beta{\hat{\tau}}_{\textrm{init}} - \nu & \beta{\hat{\tau}}_{\textrm{init}} - \nu
      \end{array} \right).
\end{equation}

One of the main results of~\cite{Rangan:10arXiv-GAMP},
which is an extension of the analysis in \cite{BayatiM:11},
was to demonstrate the convergence of the error performance of 
the GAMP algorithm to the SE equations.
Specifically, these works consider the case where
$\Abf$ is an i.i.d.\ Gaussian matrix, $\xbf$ is i.i.d.\ with a prior 
$p_X$ and $m,n \rightarrow \infty$
with $n/m \rightarrow \beta$. Then, under some further
technical conditions, it is shown that for
any fixed iteration number $t$,
the empirical joint distribution
of the components $(x_j,\xhat^t_j)$ of the unknown vector
$\xbf$ and its estimate $\xbfhat^t$ converges to a 
simple scalar equivalent model parameterized by the outputs
of the SE equations.  From the scalar equivalent
model, one can compute any asymptotic componentwise 
performance metric.  It can be shown, in particular, that 
the asymptotic MSE is 
given simply by $\bar{\tau}_t$.  That is,
\begin{equation} \label{eq:tauMse}
    \bar{\tau}_t = \lim_{n \arr \infty}
    \frac{1}{n} \sum_{j=1}^n |x_j-\xhat^t_j|^2 = 
        \lim_{n \arr \infty} \frac{1}{n} \|\xbf-\xbfhat^t\|^2.
\end{equation}
Thus, $\bar{\tau}_t$ can be used as a metric for the design
and analysis of the quantizer, although other non-squared
error distortions could also be considered.
Details are provided in \cite{Rangan:10arXiv-GAMP}.

The analysis in \cite{BayatiM:11} and \cite{Rangan:10arXiv-GAMP}
are for large i.i.d.\ Gaussian matrices.  For certain large
sparse random matrices, results in \cite{GuoW:07} and \cite{Rangan:10-CISS} show that the same SE equation holds
and, in fact, additionally 
provide testable conditions under which GAMP is provably optimal.
Specifically, it is shown that the
SE recursion in (\ref{Equ:SE:SERecursion})
always admits at least one fixed point.  As
$t \rightarrow \infty$ the recursion decreases monotonically to its
largest fixed point and, 
 if the SE admits a unique fixed point,
then GAMP is asymptotically mean-square optimal.

Thus, despite the fact that the prior on $\xbf$ may
be non-Gaussian and the quantizer function $\Q(\cdot)$ is
 nonlinear, one can precisely characterize the exact
 asymptotic behavior
 of GAMP at least for large random transforms.

\section{Quantizer Optimization}
\label{Sec:Optimization}
Ordinarily, quantizer designs depend on the distribution of
the quantizer input, with an implicit aim of minimizing the
MSE between the quantizer input and output.
Often, only uniform quantizers are considered, in which case the
``design'' is to choose the loading factor of the quantizer.
When quantized data is used as an input to a nonlinear function,
overall system performance may be improved by adjusting the
quantizer designs appropriately~\cite{MisraGV:11}.
In the present setting, conventional quantizer design minimizes
$m^{-1}\E[\|\zbf - \Op{Q}(\zbf)\|^2]$, but minimizing
$n^{-1}\E[\|\xbf - \xbfhat\|^2]$ is desired instead.

The SE description of GAMP performance facilitates the desired optimization.
By modeling the quantizer as part of the channel and working out the
resulting equations for GAMP and SE, we can make use of the
convergence result \eqref{eq:tauMse} 
to recast our optimization problem to
\begin{equation}
  \label{Equ:Quantization:SeOptimal}
   \Op{Q}^{\textrm{SE}} = \argmin_{\Op{Q}} \left\{\lim_{t \rightarrow \infty} \bar{\tau}_{t}\right\},
\end{equation}
where minimization is done over all $K$-level regular scalar quantizers.
Based on \eqref{eq:tauMse}, the optimization is equivalent
to finding the quantizer that minimizes the asymptotic MSE\@.
In the optimization \eqref{Equ:Quantization:SeOptimal},
we have considered the limit in the iterations, $t \arr \infty$.
One can also consider the optimization with a finite $t$,
although our simulations exhibit close to the limiting 
performance with a relatively small number of iterations.

It is important to note that the SE recursion behaves well under
quantizer optimization.
This is due to the fact that SE is independent of actual output levels
and small changes in the quantizer boundaries result in only minor change
in the recursion (see (\ref{Equ:RBP:Eout})).
Although closed-form expressions for the derivatives of $\bar{\tau}_{t}$
for large $t$'s are difficult to obtain, we can approximate them by using
finite difference methods.
Finally, the recursion itself is fast to evaluate, which makes the scheme in
(\ref{Equ:Quantization:SeOptimal}) practically realizable under standard
optimization methods.

\section{Experimental Results}
\label{Sec:Simulations}

\subsection{Overcomplete Expansions}
Consider overcomplete expansion of $\xbf$ as discussed in
Section~\ref{Sec:Intro:QFE}.  We generate the signal $\xbf$
with i.i.d.\ elements from the standard Gaussian distribution
$x_j \sim \mathcal{N}( 0, 1 )$.
We form the measurement matrix $\Abf$ from i.i.d.\ zero-mean
Gaussian random variables. To concentrate on the degradation due to
quantization we assume noiseless measurement model (\ref{eq:generalModel});
i.e., $\sigma^2 = 0$ in (\ref{eq:modelWithNoise}).

\begin{figure}
  \begin{center}
    \includegraphics[width=8.5cm]{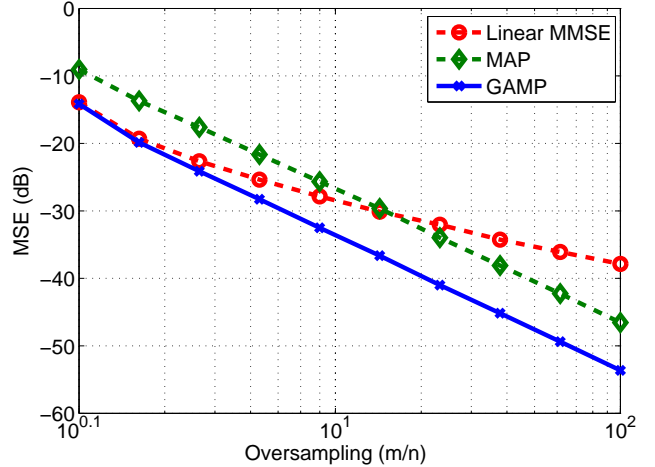}
    \caption{Performance comparison for oversampled observation of a
       jointly Gaussian signal vector (no sparsity).
       GAMP outperforms linear MMSE and MAP estimators.} 
    \label{Fig:OvercompleteMse}
  \end{center}
\end{figure}

Figure~\ref{Fig:OvercompleteMse} presents squared-error performance of three
estimation algorithms while varying the oversampling ratio $m/n$
and holding $n = 100$. To generate the plot we considered estimation from
measurements discretized by a $16$-level regular uniform quantizer.
We set the granular region of the quantizer to $[-3\sigma_z,3\sigma_z]$, where
$\sigma_z^2 = n/m$ is the variance of the measurements.
For each value of $m/n$, 200 random realizations of the problem were generated; 
the curves show the median-squared error performance over these 200 Monte Carlo trials.
We compare error performance of GAMP against two other common
reconstruction methods: linear MMSE and maximum a posteriori probability (MAP).
The MAP estimator was implemented using quadratic programming (QP).

The MAP estimation is type of \emph{consistent reconstruction}
method proposed in~\cite{ThaoV:94,ThaoV:94b,GoyalVT:98,RanganG:01,Cvetkovic:03,
BenedettoPY:06,BodmannP:07,BodmannL:08,Powell:10};
since the prior is a decreasing function of $\|\xbf\|$,
the MAP estimate $\xbfhat$ is the vector consistent with
$\Q(\Abf\xbfhat)$ of minimum Euclidean norm.
In the earlier works, it is argued that consistent
reconstruction methods offer improved performance
over linear estimation,
particularly at high oversampling factors.
We see in Figure~\ref{Fig:OvercompleteMse} that MAP
estimation does indeed outperform linear MMSE at high oversampling.
However, GAMP offers significantly better performance 
than both LMMSE and MAP,
with more than 5~dB improvement for many values of $m/n$.
In particular, this reinforces that MAP is suboptimal because
it finds a corner of the consistent set, rather than the centroid.
Moreover, the GAMP method is actually computationally simpler
than MAP, which requires the solution to a quadratic program.

\begin{figure}
\vspace{-0.2cm}
  \begin{center}
    \includegraphics[width=8.5cm]{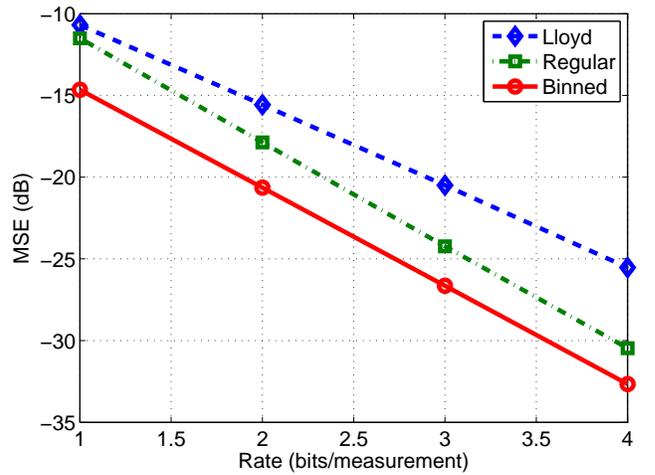}
    \caption{Performance comparison of GAMP with optimal uniform quantizers under Gaussian prior for regular and binned quantizers.} 
    \label{Fig:RegBinnedMseOversampled}
  \end{center}
\end{figure}

With Figure~\ref{Fig:RegBinnedMseOversampled} we turn to a comparison among
quantizers, all with GAMP reconstruction, $n=100$, $m=200$, and
$\xbf$ and $\Abf$ distributed as above.
To demonstrate the improvement in rate--distortion performance that is
possible with non-regular quantizers,
we consider simple \emph{uniform modulo} quantizers
\begin{equation}
\label{Equ:ModuloQuantizer}
\Op{Q}\left(z\right) = \left\lfloor \frac{z}{\Delta} \right\rfloor \mod N,
\end{equation}
where $\Delta$ is the size of the quantization cells.
These quantizers map the entire real line $\R$ to the set $\{0,\,1,\,\ldots,\,N-1\}$ in a periodic fashion.

We compare three types of quantizers:
those optimized for MSE of the measurements
(\emph{not} the overall reconstruction MSE)
using Lloyd's algorithm~\cite{GrayN:98},
regular uniform quantizers with loading factors
optimized for reconstruction MSE using SE analysis,
and
(non-regular) uniform modulo quantizers with $\Delta$
optimized for reconstruction MSE using SE analysis.
The last two quantizers were obtained by solving \eqref{Equ:Quantization:SeOptimal} via the standard SQP method found in MATLAB\@.
The uniform modulo quantizer achieves the best rate--distortion performance,
while the performance of the quantizer designed with Lloyd's algorithm is comparatively poor.
The stark non-optimality of the latter is due to the fact that it optimizes the MSE only between quantizer inputs and outputs, ignoring the nonlinear estimation algorithm following the quantizer.

It is important to point out that, without methods such as GAMP,
 estimation  with
a modulo quantizer such as \eqref{Equ:ModuloQuantizer} is not
even computationally possible in works such as
\cite{ThaoV:94,ThaoV:94b,GoyalVT:98,RanganG:01,Cvetkovic:03,
BenedettoPY:06,BodmannP:07,BodmannL:08,Powell:10},
since the consistent set is non-convex and
consists of a disjoint union of convex sets.
Beyond the performance improvements, we believe that GAMP
provides the first computationally-tractable and systematic
method for such non-convex quantization reconstruction
problems.

\subsection{Compressive Sensing with Quantized Measurements}
We next consider estimation of
an $n$-dimensional sparse signal $\xbf$
from $m < n$ random measurements---a problem considered
in quantized compressed sensing 
\cite{ZymnisBC:10,JacquesHF:11,LaskaBDB:11}.
We assume that the signal $\xbf$
is generated with i.i.d.\ elements from the Gauss--Bernoulli distribution
\begin{equation}
   x_j \sim \left\{
      \begin{array}{ll}
	\mathcal{N}\left( 0, 1/\rho \right), & \mbox{with probability $\rho$};\\
	                   0,              & \mbox{with probability $1 - \rho$},
      \end{array}
\right.
\end{equation}
where $\rho$ is the sparsity ratio that represents the average fraction
of nonzero components of $\mathbf{x}$.
In the following experiments we assume $\rho = 1/32$.
Similarly to the overcomplete case, we form the measurement matrix $\Abf$
from i.i.d.\ Gaussian random variables and we assume no additive noise ($\sigma^2 = 0$ in (\ref{eq:modelWithNoise})).

\begin{figure}
  \begin{center}
    \includegraphics[width=8.5cm]{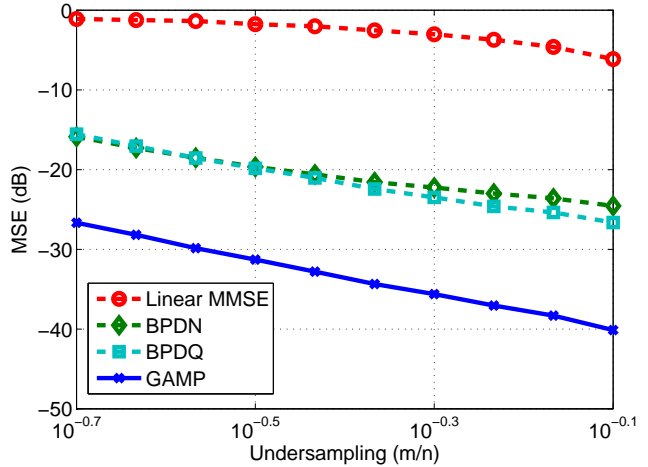}
    \caption{Performance comparison of GAMP with LMMSE, BPDN, and BPDQ (with moment $p = 4$) for estimation from compressive measurements.} 
    \label{Fig:UndercompleteMse}
  \end{center}
\end{figure}

Figure~\ref{Fig:UndercompleteMse} compares MSE performance of GAMP
with three other standard reconstruction methods. In particular, we consider
linear MMSE and the Basis Pursuit DeNoise (BPDN) program~\cite{CandesRT:06-CPAM} 
\begin{equation*}
  \xbfhat = \argmin_{\xbf \in \R^n } \| \xbf \|_{1} \textrm{ s.t. } \| \ybf- \Abf\xbf \|_p \leq \epsilon,
\end{equation*}
where $p = 2$ and $\epsilon \in \R_{+}$ is the parameter representing the noise power. In the same figure, we additionally plot
the error performance of the Basis Pursuit DeQuantizer (BPDQ) of moment $p$, proposed in~\cite{JacquesHF:11}, which solves
the problem above for $p \geq 2$. It has been argued in~\cite{JacquesHF:11} that BPDQ offers better error performance compared to the standard BPDN as the number of samples $m$ increases with respect to the sparsity $k$ of the signal $\xbf$.

We obtain the curves by varying
the ratio $m/n$ and holding $n = 1024$.
We perform estimation from measurements obtained from a 
$16$-level regular uniform quantizer with granular region of length $2\|\Abf\xbf\|_{\infty}$ centered at the origin.

The figure plots the median of the squared error from 1000 Monte Carlo trials
for each value of $m/n$.
For basis pursuit methods we optimize the parameter $\epsilon$
for the best squared error performance;
in practice this oracle-aided performance would not be achieved.
The top curve (worst performance) is for linear MMSE
estimation; and middle curves are for the basis pursuit estimators BPDN and BPDQ with moment $p = 4$. As expected, BPDQ
achieves a notable $2$ dB reduction in MSE compared to BPDN for high values of $m$, however GAMP significantly
outperforms both methods over the whole range of $m/n$.

\begin{figure}
  \begin{center}
    \includegraphics[width=8.5cm]{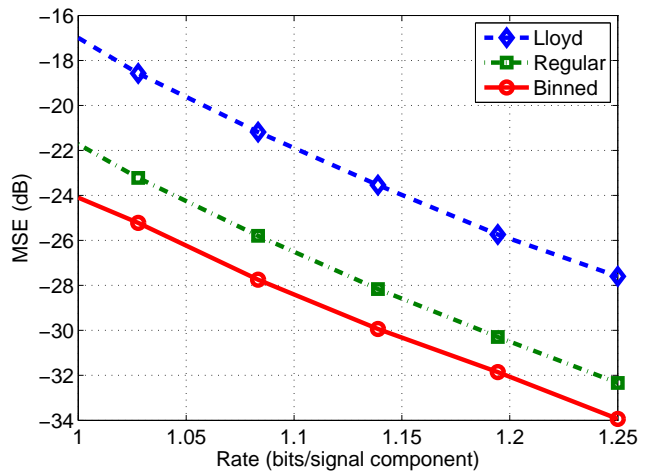}
    \caption{Performance comparison of GAMP with optimal uniform quantizers under Gauss-Bernoulli prior for regular and binned quantizers.} 
    \label{Fig:RegBinnedMse}
  \end{center}
\end{figure}

In Figure~\ref{Fig:RegBinnedMse}, we compare the performance of GAMP under three quantizers consider before: 
those optimized for MSE of the measurements
using Lloyd's algorithm, and
regular and non-regular quantizers optimized
for reconstruction MSE using SE analysis.
We assume the same $\xbf$ and $\Abf$ distributions as above.
We plot MSE of the reconstruction against the rate measured 
in bits per component of $\xbf$. For each rate and for each quantizer, we vary the ratio $m/n$ for the best possible performance.
We see that, in comparison to regular quantizers, binned quantizers with GAMP
estimation achieve much lower distortions for the same rates.
This indicates that binning can be an effective strategy to favorably
shift rate--distortion performance of the estimation.

\section{Conclusions}
\label{Sec:Conclusions}
We have presented generalized approximate message passing as an effective and
efficient algorithm for estimation from quantized linear measurements.
The GAMP methodology is general, allowing essentially
arbitrary priors and quantization functions.
In particular, GAMP is the first tractable and effective method for
high-dimensional estimation problems
involving non-regular scalar quantization.
In addition, the algorithm is computationally extremely simple and,
in the case of large random transforms, admits a precise
performance characterization using a state evolution analysis.

The problem formulation is Bayesian, with an i.i.d.\ prior over the
components of the signal of interest $\xbf$; the prior may or may not
induce sparsity of $\xbf$.  Also, the number of measurements may be
more or less than the dimension of $\xbf$, and the quantizers applied
to the linear measurements may be regular or not.
Experiments show significant performance improvement
over traditional reconstruction schemes,
some of which have higher computational complexity.
Moreover, using extensions of GAMP such as hybrid approximate message passing
\cite{RanganFGS:11arXiv}, one may also in the future 
be able to consider quantization of more general classes of signals
described by general graphical models. 
MATLAB code for experiments with GAMP is available online~\cite{Rangan:GAMP-code}.

Despite the improvements demonstrated here, we are not advocating
quantized linear expansions as a compression technique---for the oversampled
case or the undersampled sparse case; thus, comparisons to rate--distortion
bounds would obscure the contribution.
For regular quantizers and some fixed oversampling $\beta = m/n > 1$,
the MSE decay with increasing rate is $\sim 2^{-2R/\beta}$, worse than
the $\sim 2^{-2R}$ distortion--rate bound.
For a discussion of achieving exponential decay of MSE with increasing
oversampling, while the quantization step size is held constant,
see~\cite{CvetkovicV:98-IT}.
For the undersampled sparse case, \cite{GoyalFR:08} discusses the difficulty
of recovering the support from quantized samples and the consequent difficulty
of obtaining near-optimal rate--distortion performance.
Performance loss rooted in the use of a random transformation $\Abf$
is discussed in~\cite{FletcherRG:07a}.

\bibliographystyle{IEEEtran}
\bibliography{bibl}

\end{document}